\RequirePackage{rotating}

\documentclass{jfm}
\usepackage{graphicx}
\usepackage{epstopdf, epsfig}
\usepackage{graphicx}
\usepackage{rotating}
\usepackage{amsmath}
\usepackage{color}
\usepackage{etex}
\reserveinserts{18}
\usepackage{morefloats}
\usepackage{enumitem}
\usepackage{xparse}
\usepackage{graphicx,caption,varwidth}
\usepackage[draft]{changes}
\makeatletter
\newcommand*{\rom}[1]{\expandafter\@slowromancap\romannumeral #1@}
\makeatother

\shorttitle{Bistability of bidirectional flow}
\shortauthor{Suckale et al.}

\title{Bistability of buoyancy-driven exchange flows in vertical tubes}

\author{Jenny Suckale\aff{1}
  \corresp{\email{jsuckale@stanford.edu}},
  Zhipeng Qin\aff{1},
  Davide Picchi\aff{2},
  Tobias Keller\aff{1},
  Ilenia Battiato\aff{2}}

\affiliation{\aff{1}Department of Geophysics, Stanford University, Stanford, California, USA.
\aff{2}Department of Energy Resources Engineering, Stanford University, Stanford, California, USA.}

\begin{document}

\maketitle

\begin{abstract}
Buoyancy-driven exchange flows are common to a variety of natural and engineering systems ranging from persistently active volcanoes to counterflows in oceanic straits. Experiments of exchange flows in closed vertical tubes have been used as surrogates to elucidate the basic features of such flows. The resulting data have historically been analyzed and interpreted through core-annular flow solutions, the most common flow configuration at finite viscosity contrasts. These models have been successful in fitting experimental data, but less effective at explaining the variability observed in natural systems. In this paper, we formulate a core-annular solution to the classical problem of buoyancy-driven exchange flows in vertical tubes. The model posits the existence of two mathematically valid solutions, i.e. thin- and thick-core solutions. The theoretical existence of two solutions, however, does not necessarily imply that the system is bistable in the sense that flow switching may occur. Using direct numerical simulations, we test the hypothesis that core-annular flow in vertical tubes is bistable, which implies that the realized flow field is not uniquely defined by the material parameters of the flow. Our numerical experiments, which fully predict experimental data without fitting parameters, demonstrate that buoyancy-driven exchange flows are indeed inherently bistable systems. This finding is consistent with previous experimental data, but in contrast to the underlying hypothesis of previous analytical models that the solution is unique and can be identified by maximizing the flux or extremizing the dissipation in the system. These results have important implications for data interpretation by analytical models, and may also have relevant ramifications for understanding volcanic degassing.

%We study buoyancy-driven exchange flow at low Reynolds number in vertical tubes. We build on existing laboratory experiments to make progress towards our fundamental understanding and ability to quantify exchange flow in vertical tubes. We reproduce these experiments virtually through direct numerical simulations in two dimensions.
%\deleted{Our simulations do not entail any tunable parameters and hence provide a direct representation of the flow dynamics in the laboratory on a computer.} 
%We then complement our simulations with a simple analytical model to make progress towards an analytical understanding of the observed flow patterns that does not require high performance computing. Our analytical model\deleted{s} posits the existence of two mathematically viable solutions for core-annular flow, the most common flow configuration at finite viscosity contrasts. Yet, only one of these solutions is observed in closed-system experiments. Our simulations suggest not only that both solutions may arise in open systems at intermediate viscosity contrasts between upwelling and downwelling melt, but also that the boundary conditions could trigger a dynamic switch between these solutions. This finding indicates that the flow is bistable. This finding is supported by laboratory experiments in more complex geometries that allow for flux in and out of the tube.
%Our study is motivated by the need to understanding degassing in persistently degassing volcanoes, but our insights are not limited to this particular application.
\end{abstract}


\section{Introduction} \label{sec:intro}
Buoyancy-driven, bidirectional flow in channels or tubes is relevant to many natural and industrial processes. Examples include the joint pumping of viscous oil and water through pipelines
\citep[e.g.,][]{Joseph1997}, the flow of cement into drilling mud in wellbores \citep[e.g.,][]{frigaard1998uniaxial}, the counterflow of currents with different densities in oceanic straits \citep[e.g.,][]{dalziel1992maximal}, and magma circulation in the conduits of persistently degassing volcanoes \citep[e.g.,][]{Stevenson1998}. If the fluids are immiscible, the emerging flow is linearly unstable \citep[e.g.,][]{Joseph1997} and significant effort has been devoted to identifying the pertinent flow regimes and their relative stabilities \citep[e.g.,][]{barnea1987unified,Brauner1998}.

Here, we are interested in understanding the exchange flow of two Newtonian fluids in a vertical tube at low Reynolds number. Our work is motivated primarily by a need to better understand degassing processes in persistently active volcanoes, the most common form of volcanism on Earth. Many persistently degassing volcanoes have been active for periods exceeding the historical record. While erupting comparatively small amounts of lava, they continuously emit copious amounts of volatiles and thermal energy, suggesting that the majority of magma must be recycled back into the plumbing system after decoupling from the gas phase near the surface \citep[e.g.,][]{Francis1993}. At steady state, 
%JS: Minor thing, I think the correct use is "at steady state" where "steady state" is a compound noun, vs "in a/the steady state" where "steady" is an adjective modifying "a/the state". Propagated this correction throughout.
the ascent of gas-rich, buoyant magma in the conduit would therefore approximately balance the simultaneous descent of gas-poor, dense magma, which would result in exchange flow in the volcanic conduit \citep{Kazahaya1994,Stevenson1998,burton2007role,Witham2011}.
%JS: Introduced conditional tense in this sentence to highlight that, at this stage, we are dealing with a suggestion, not a proven outcome.

Our work builds on existing laboratory experiments of exchange flow in closed vertical tubes by \citet{Stevenson1998}. We focus on this specific set of analogue experiments, as their relatively simple geometry provides a suitable starting point towards understanding more complex geometries \citep{Huppert2007,Beckett2011,palma2011}. We analyze the flow behavior observed in the laboratory by using direct numerical simulations to virtually reproduce the original experiments. Our numerical model does not require closures such as drag forces, interface stresses or rise speeds \citep{Suckale2010a, Suckale2010, Qin2017}. Instead, these flow properties emerge self-consistently from the computation, similar to analogue experiments where the flow dynamics emerge directly from the experimental setup and the materials used. 
%This approach has two advantages. First, our virtual reproductions of the actual experiments enable us to quantify velocities, stresses and other flow variables at all times and locations in the flow field. Second, our simulations are not limited to scales or flow conditions that are achievable in a laboratory setting, which will allow a generalized understanding of the flow behavior observed in the experiments. 

The most commonly observed flow regime of exchange flows in closed vertical tubes has a core-annular geometry, where one fluid flowing in the center of the tube (core) is surrounded by a film of the other fluid wetting the tube walls (annulus), as schematically illustrated in figure~\ref{fig:1} \citep{bai1992lubricated,petitjeans1996miscible,Stevenson1998,Scoffoni2001,Kuang2003}.
The result that core-annular flow appears to be the dominant flow regime for bidirectional flow in the presence of substantial viscosity contrasts may seem counter-intuitive. Core-annular flow is prone to instability, as has been shown through linearized stability analysis by numerous studies reviewed in \citet{Joseph1997}. However, high viscosity contrasts and surface tensions may suppress the growth rate of interface instabilities sufficiently to stabilize core-annular flow \citep[e.g.,][]{Hickox1971}. If the two fluids are miscible, even a small degree of mixing at the interface can further stabilize the flow configuration over a wide range of conditions \citep{meiburg2004density}. 

The prominence of the core-annular regime in exchange flow is fortuitous, because core-annular flow is amenable to analytical analysis \citep[e.g.,][]{russell1959effect,Kazahaya1994,Ullmann2004,Huppert2007}. Here, we build on \citet{Ullmann2004} to derive a simple analytical model of steady core-annular flow in a vertical tube. 
%the volcanic context. While our analytical approach does not encapsulate the full complexity of the flow behavior observed in the analogue and virtual experiments, it provides the basis for a simple conduit flow model that provides predictive and explanatory power at minimal computational cost. 
Our analysis generalizes the work of \citet{Kazahaya1994} by allowing for a mobile interface between the ascending and descending fluids. This generalization resolves the mismatch between theoretical predictions and laboratory observations reported by \citet{Stevenson1998}. Our analytical model improves upon the formulation of \citet{Huppert2007} by eliminating the approximate representation of pressure and employing an alternative non-dimensionalization. The model predicts the existence of two distinct solutions at steady state and our numerical simulations confirm that both solutions can be realized in practice. This finding is consistent with laboratory experiments in more complex geometries \citep[e.g.,][]{Beckett2011}, which allow for flow in and out of an open vertical tube. We conclude that exchange flow in vertical tubes is a bistable system---an aspect of core-annular flow that could have important implications for understanding magma circulation in volcanic conduits.

%------------------------------------------------

\section{Model Description} \label{sec:model}
%To understand bidirectional conduit flow 
%and the decompression histories experienced by magma undergoing such flow, 
We combine two distinct and complementary model components: an analytical model of core-annular flow at steady state, and a direct numerical model of exchange flow in two dimensions (2D). An advantage of the analytical model is its applicability in both 2D and 3D, which is important for a detailed comparison of 3D laboratory experiments with numerical simulations in 2D. An advantage of the numerical model is the ability to capture the various flow regimes that arise for both the miscible fluids employed in the laboratory experiments and the immiscible fluids assumed in the analytical model. By combining the two approaches, we are able to leverage the flexibility of the numerical model in concert with the simplicity and explanatory power of the analytical model.

\subsection{Analytical Model} \label{sec:analyticmodel}
\subsubsection{Derivation} \label{sec:analyticderiv}
\begin{figure}
	\centering
	\includegraphics[width=0.5\textwidth]{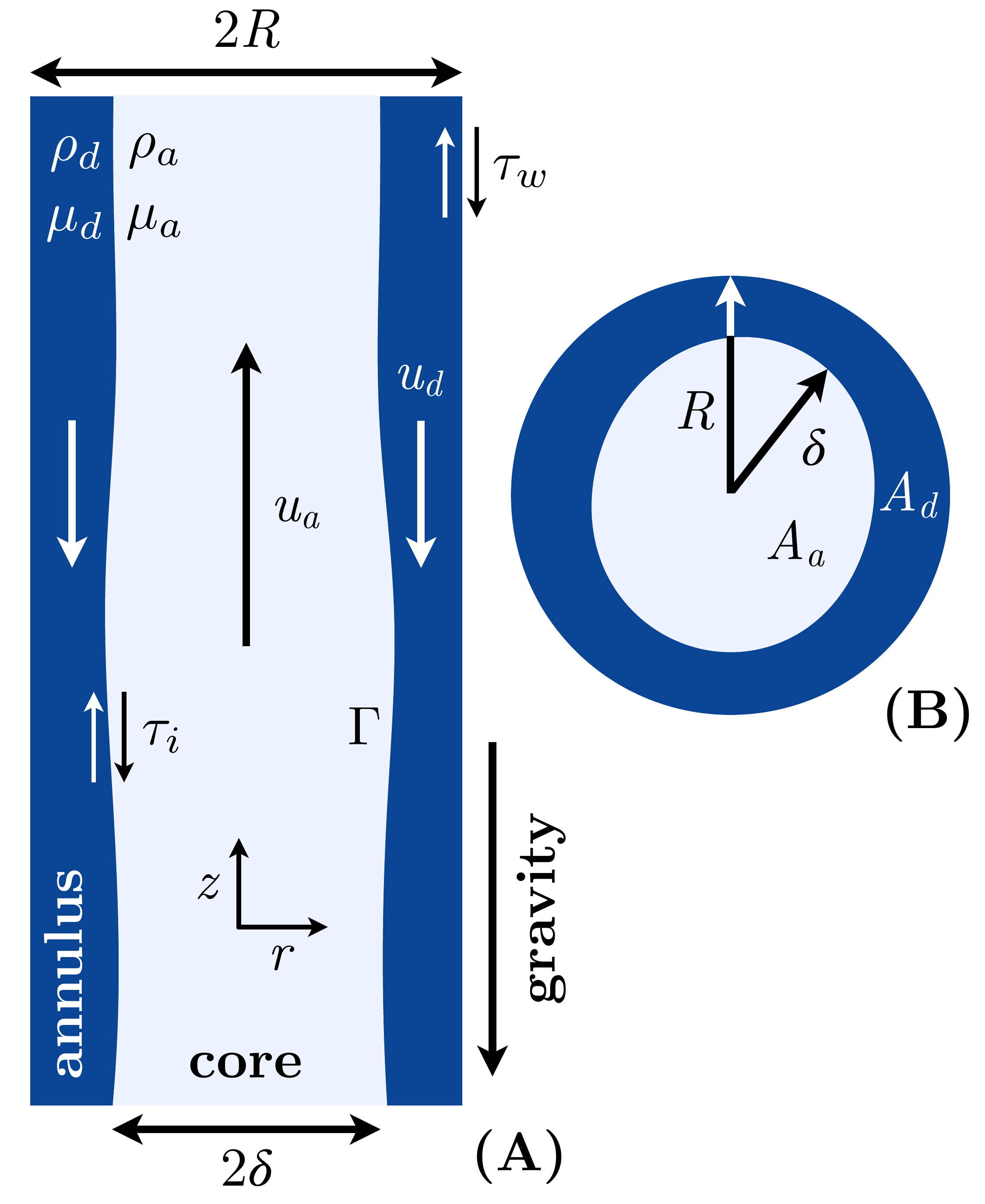}
	\caption{\small Sketch of the core-annular flow geometry and key variables used for a vertical (A) and a horizontal (B) cross section. The interface between the two fluids is depicted as wavy to highlight the unstable nature of the flow pattern.}
	\label{fig:1}
\end{figure}

%Previous efforts to understand bidirectional flow \citep[e.g.,][]{bai1992lubricated, petitjeans1996miscible,Stevenson1998,Scoffoni2001,Kuang2003} suggest that core-annular flow is the dominant regime in closed vertical tubes. We hence focus on that configuration for deriving our analytical model \citep{Ullmann2004, Huppert2007}.
%
At low Reynolds number, the  fully-developed, steady-state core-annular flow of two immiscible and incompressible fluids in a pipe of inclination $\alpha$ from the vertical direction can be described by the 1D Navier-Stokes equations along the radial coordinate  $r \in [0,R]$ (see figure\ref{fig:1}),
%\deleted{At steady state, core-annular flow does not depend on the vertical coordinate, $z$, which reduces the problem to a 1D force balance in each fluid in the across-flow direction, along the radial coordinate $r = [0,R]$ (see Fig.~\ref{fig:1}),}
\begin{subequations}\label{eq:A}
\begin{eqnarray}
\mu_d\frac{1}{r}\frac{\mathrm d}{\mathrm d r} \left(  r \frac{\mathrm d u_d}{\mathrm d r} \right)&=&\frac{\mathrm dp}{\mathrm dz}+ \rho_d g \cos \alpha,  \quad r \in [\delta,R],\label{eq:A1} \\
{\mu_a}\frac{1}{r}\frac{\mathrm d}{\mathrm d r} \left(  r \frac{\mathrm d u_a}{\mathrm d r} \right)&=&\frac{\mathrm dp}{\mathrm dz} +\rho_a g \cos \alpha,  \quad r \in [0,\delta],\label{eq:A2}
\end{eqnarray}
\end{subequations}
where the subscripts $(\cdot)_d$ and $(\cdot)_a$ denote the descending and the ascending fluids, respectively, $R$ is the pipe radius and $\delta$ represents the unknown location of the interface between the ascending and descending fluids. Unlike in the core-annular model proposed by \citet{Huppert2007} where the form of the pressure drop is postulated \citep[Eq. (4.4)]{Huppert2007}, $\mathrm dp/\mathrm dz$ is an unknown constant to be determined here. The boundary conditions are no-slip at the tube wall,
\begin{eqnarray}
u_d(R)=0 \, ,
\end{eqnarray}
vanishing stress at the symmetry line in the tube center,
\begin{eqnarray}
\frac{\mathrm d u_a}{\mathrm d r}(0)=0,
\end{eqnarray}
and continuity of velocity and shear stress across the fluid-fluid interface,
\begin{eqnarray}
u_d(\delta)-u_a(\delta)&=&0, \\
\mu_a\frac{\mathrm d u_a}{\mathrm d r}(\delta)-{\mu_d}\frac{\mathrm d u_d}{\mathrm d r}(\delta)&=&0.
\end{eqnarray}
Instead of a phase flow-rate scaling \citep{Ullmann2004}, we define the non-dimensional quantities % 
\begin{align}\label{eq:dimensionless}
\hat{r}=\dfrac{r}{R}, \quad \hat{\delta}=\dfrac{\delta}{R}, \quad \hat{u}=\dfrac{u}{U},
\end{align}
where  $U:=\Delta \rho g R^2 / \mu_d$ is the viscous rise speed. Substituting \eqref{eq:dimensionless} into \eqref{eq:A} and dropping all hats yields the dimensionless equations,
\begin{subequations}\label{eq:dimensionlessA}
\begin{eqnarray}
\frac{1}{r}\frac{\mathrm d}{\mathrm d r} \left(  r \frac{\mathrm d u_d}{\partial r} \right)&=&P,  \quad r \in [\delta,1] \label{eq:A1a} \\
\frac{1}{{\mathrm{M}}} \frac{1}{ r}\frac{\mathrm d}{\mathrm d r} \left(  r \frac{\mathrm d u_a}{\mathrm d r} \right)&=&P-\cos \alpha,  \quad r \in [0,\delta] \label{eq:A2a}
\end{eqnarray}
\end{subequations}
where  $P=(\mathrm dp/\mathrm dz+\rho_d g \cos \alpha)/(g\Delta \rho)$ is the non-dimensional pressure drop driving the descending phase, M$=\mu_d/\mu_a$ is the viscosity ratio, and $\Delta \rho=\rho_d-\rho_a$ is the density difference. We integrate equations \eqref{eq:dimensionlessA}, while accounting for appropriately nondimensionalized boundary conditions, to obtain 
\begin{subequations}\label{solution}
\begin{eqnarray}
\label{eq:A3}
u_d(r)&=&\frac{P}{4} \left(r^2-1\right)-\frac{{\delta}^2}{2}\cos \alpha \log r, \quad r \in [\delta,1] \\ 
u_a(r)&=&{\mathrm{M}}\frac{P-\cos \alpha}{4 }\left( r^2-{\delta}^2\right)+    u_i, \quad r \in [0,\delta] \label{eq:A4}
\end{eqnarray}
\end{subequations}
where $u_i=u_d(\delta)=u_a(\delta)$ is the vertical flow speed at the interface given by 
\begin{eqnarray}
\label{eq:ui}
u_i=\frac{P}{4 }\left({\delta}^2-1\right)- \frac{{\delta}^2}{2}\cos \alpha \log \delta.
\end{eqnarray}

The ascending flux in a closed tube with incompressible fluids must exactly balance the descending flux,
\begin{eqnarray}\label{eq:A5}
- \int _{{\delta}}^1 2\pi r u_d(r) dr=\int _0 ^{{\delta}} 2\pi r u_a(r) dr={\mathrm{Te}} \, ,
\end{eqnarray}
where Te is the dimensionless flux or Transport number \citep[e.g.,][]{Ullmann2004,Huppert2007}. The Transport number is related, but not identical, to the Poiseuille number, Ps, defined in \citet{Stevenson1998},
\begin{eqnarray}
{\mathrm{Ps}} = \frac{u \mu_d}{g \Delta \rho R^2} = \frac{u}{U}  \, ,
\end{eqnarray}
where $u$ is the (dimensional) terminal rise speed. The Ps number therefore represents the non-dimensional rise speed, whereas the Te number captures the non-dimensional flux and hence
\begin{eqnarray}
\label{eq:Te-Ps}
{\mathrm{Te}} = \frac{{\mathrm{Ps}} }{\pi \delta^2}  \, .
\end{eqnarray}
We focus on Te in our analysis, becuase the flux is balanced between ascending and descending fluids and does not hinge on selecting either a characteristic rise speed from the spatially variable function, $u_a(r)$, or a core radius, $\delta$.  To fulfill the constraint of no-net flux in the tube, we substitute \eqref{solution} into~\eqref{eq:A5} and solve for the driving force $P$ that depends only on the dimensionless parameters of the problem (i.e., ${\delta}$, $\mathrm{M}$, $\alpha$),
\begin{eqnarray}\label{eq:A6}
P=   {\delta}^2   \frac{2 {({\delta}^2-1)}-  {\mathrm{M}}{\delta}^2 }{({\delta}^4-1)-{\mathrm{M}}{\delta}^4} \cos \alpha.
\end{eqnarray}
We can now express Te as a function of $P$ and $\delta$
\begin{eqnarray}\label{eq:A7}
{\mathrm{Te}}= 2 \pi \left[  \frac{P}{16}({\delta}^2-1)^2+\frac{{\delta}^2}{8} \cos \alpha ({\delta}^2-1-2{\delta}^2\log {\delta})		 \right].
%DP: please double check this equation, where I corrected a bug in the second \frac.
\end{eqnarray}
We note that once $P$ and $\delta$ are defined, Te is uniquely specified. The opposite, however, is not true since Te is a quadratic function of $\delta$. For any Te below the maximum possible flux or flooding point, there are mathematically valid solutions. Despite differences in some of its technical ingredients, the model by \citet{Huppert2007} also admits multiple solutions, but the authors implicitly impose uniqueness by maximizing the flux or, alternatively, extremizing the dissipation in the system. In \S\ref{sec:results}, we numerically explore the validity of this uniqueness hypothesis and investigate whether both solutions may be realized.

\subsubsection{Special Case: 2D Geometry} \label{sec:analytic2D}
The analytical model we have now laid out for a general tube geometry in 3D can also be formulated in 2D. The latter is useful to compare our 2D numerical simulations to the 3D laboratory experiments. In the 2D case, the non-dimensional vertical speed in the two fluids is 
\begin{eqnarray}
\label{eq:ud_2D}
u_d(y)&=&\frac{P}{2} \left(y^2-1\right)-	\delta \cos \alpha (y-1), \quad y \in [\delta,1] \\ 
u_a(y)&=&{\mathrm{M}}\frac{P-\cos \alpha}{2 }\left( y^2-{\delta}^2\right)+    \frac{P}{2} \left( {\delta}^2-1\right) - \delta \cos \alpha \left( {\delta}-1\right), \quad y \in [0, \delta] \label{eq:ua_2D}
\end{eqnarray}
where $0\leq y \leq 1$ is the Cartesian direction perpendicular to the side boundaries and $\delta$ is the half-thickness of the ascending fluid. We will later use these expressions as boundary conditions to force bidirectional flow in our 2D numerical model (see \S\ref{sec:results-general}).

\subsection{Numerical Model} \label{sec:numerics}
\subsubsection{Governing Equations} \label{sec:governingeqs}
Our numerical model solves for conservation of mass and momentum. We assume that both fluids are incompressible. The governing equations are hence the incompressibility condition,
\begin{eqnarray} \label{eq:GE1}
\nabla\cdot{\bf{v}}=0 \ ,
\end{eqnarray}
and the Navier-Stokes equation, 
\begin{eqnarray} \label{eq:GE2}
\rho\left(\frac{\partial{\bf{v}}}{\partial t}+({\bf{v}}\cdot\nabla){\bf{v}}\right)=-\nabla p+\nabla\cdot\left[\mu(\nabla{\bf{v}}+(\nabla{\bf{v})}^T)\right]+\rho\bf{g} \ .
\end{eqnarray}
where $\bf{v}$ is the velocity, $p$ the pressure, $\bf{g}$ the gravitational acceleration, $\rho$ the density, and $\mu$ the dynamic viscosity. We assume that both fluids abide by a Newtonian rheology. 

Density and viscosity vary in space and time to reflect the different properties of the two fluids. We consider the two contrasting scenarios of immiscible and miscible fluids. The immiscible case is useful for a direct comparison with the analytical solution, which is strictly valid only for two immiscible fluids. The miscible case represents the experimental setup by \citet{Stevenson1998}, which involves two miscible fluids with low diffusivity. 

\subsubsection{Immiscible fluids} \label{sec:immisciblemodel}
In the immiscible case, a sharp interface separates the two fluids. We advect the curve, $\Gamma$, which represents this interface, with the flow field according to
\begin{eqnarray}\label{eq:GE6}
{\frac{\partial{\Gamma}}{\partial t}}+({\bf{v}}\cdot\nabla)\Gamma=0 \ . 
\end{eqnarray}
The interface deforms in response to the hydrodynamic forces acting on it. The material properties change discontinuously at the interface
\begin{eqnarray}\label{eq:GE7}
\mu({\bf{x}}) = \left\{ \begin{array}{rl}
\mu_a & \mbox{in the ascending fluid}
\\
\mu_d & \mbox{in the descending fluid} \\
%\mu_s & \mbox{for} & {\bf{x}}\in \text{the solid}
\end{array}\right.  \ ,
\end{eqnarray}
and
\begin{eqnarray}\label{eq:GE8}
\rho({\bf{x}}) = \left\{ \begin{array}{rl}
\rho_a & \mbox{in the ascending fluid}
\\
\rho_d & \mbox{in the descending fluid} \\
%\mu_s & \mbox{for} & {\bf{x}}\in \text{the solid}
\end{array}\right.  \ 
\end{eqnarray}
and may entail a jump in the pressure and normal stresses at the interface:
\begin{eqnarray}\label{eq:GE9}
\left[\left(\begin{array}{rcl}
{\bf{n}} \\
{\bf{t}}_1\\
{\bf{t}}_2
\end{array}\right)\left(p\bf{I}-\boldsymbol{\tau}\right){\bf{n}}^{\text{T}}\right]=\left(\begin{array}{rcl}
\sigma\kappa\\
0\\
0
\end{array}\right) \ ,
\end{eqnarray}
Here, square brackets $[\cdot]$ denote a jump at the fluid-fluid interface, $\bf{I}$ is the identity tensor, $\boldsymbol{\tau} = \mu\left(\nabla{\bf{v}}+(\nabla{\bf{v})}^T\right)$ the viscous stress tensor, $\sigma$ the surface tension, $\kappa$ the curvature of $\Gamma$, $\bf{n}$ the unit normal vector on $\Gamma$ pointing from the ascending towards the descending fluid, and ${\bf{t}}_1$ and ${\bf{t}}_2$ the two unit tangential vectors on $\Gamma$.

\subsubsection{Miscible fluids} \label{sec:misciblemodel}
To allow for mixing, we introduce a continuous variable, $c$, which represents the concentration of the buoyant fluid. Initially, the concentration is set to unity in the buoyant fluid and zero in the heavy fluid. The concentration evolves over time due to advection and diffusion,
\begin{eqnarray} \label{eq:GE3}
\frac{\partial c}{\partial t} + {\bf{v}}\cdot\nabla c = D \nabla^2c \ ,
\end{eqnarray}
where $D$ is the diffusion coefficient. In the experiments, \citet{Stevenson1998} used miscible fluids, such as syrup, dilute syrup, glycerol and water, but they did not provide the fluid diffusivities. In the absence of detailed measurements, we use a constant diffusivity of $D=10^{-10}$ m/s$^2$ in both fluids for all simulations, a value motivated by the diffusivity measured for corn syrup in distilled water \citep{Ray2007}. 
%We have tested that the model outcome does not significantly depend on the choice of diffusivity, as long as the value is low enough for the mixing layer to remain small compared to the core thickness.

We assume that density and viscosity depend linearly on the concentration $c$, such that
\begin{eqnarray}\label{eq:GE4}
\rho = \rho_d+c \ (\rho_a-\rho_d) \
\end{eqnarray}
and
\begin{eqnarray}\label{eq:GE5}
\mu = \mu_d+c \ (\mu_a-\mu_d) \ .
\end{eqnarray}
%The subscript $(\cdot)_a$ again refers to the buoyant, ascending, and $(\cdot)_d$ to the denser, descending fluid, respectively (see figure~\ref{fig:1}). 
We note that additional complexity may arise in the vicinity of the interface if a nonlinear dependence of viscosity on concentration is assumed as is the case in some prior studies \citep[e.g.,][]{Tan1986,Goyal2006} (see online supplement for details).

\begin{figure}
	\centering
	\includegraphics[width=0.5\textwidth]{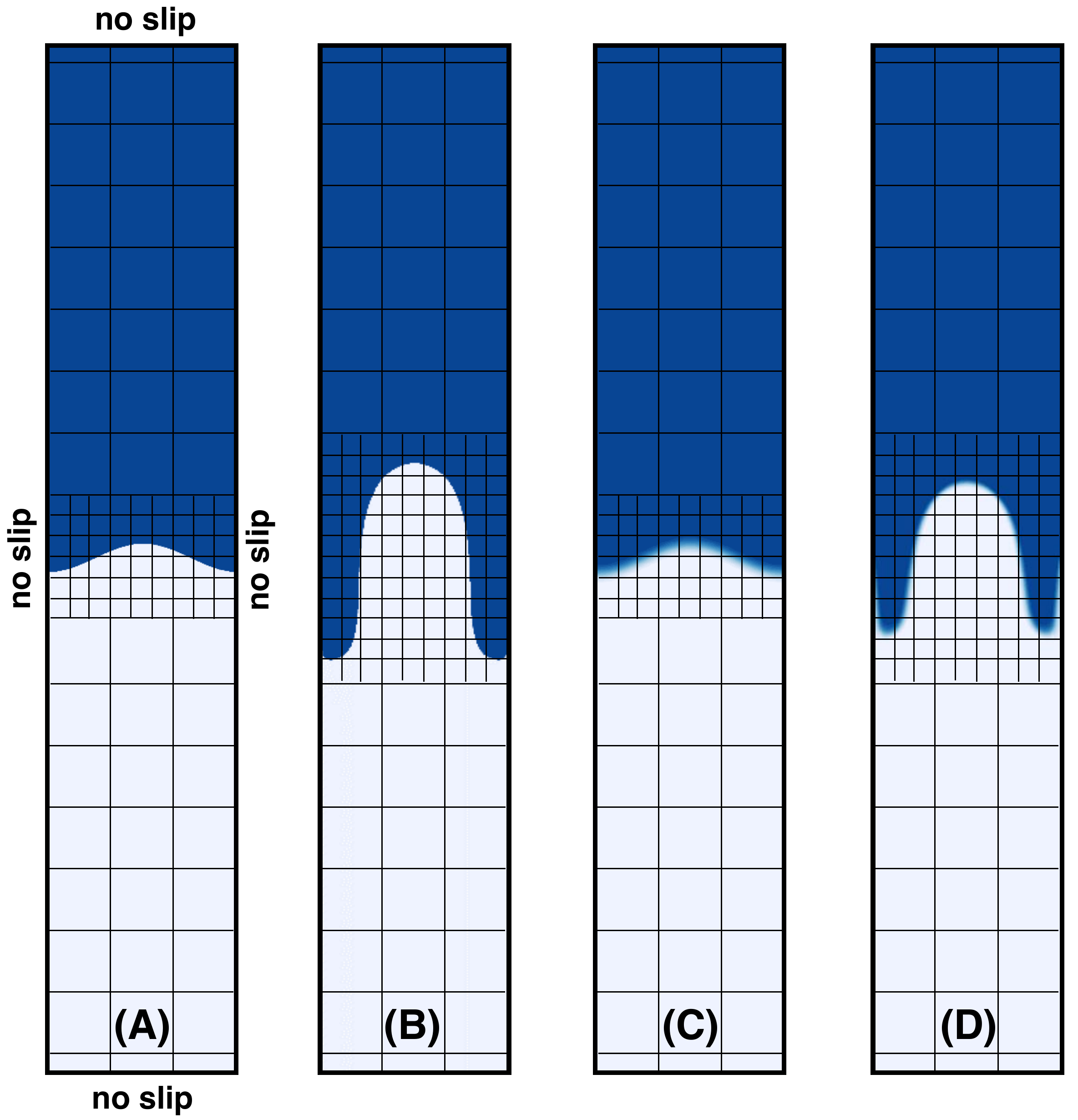}
	\caption{\small Illustration of the initial conditions, boundary conditions and adaptive grid refinement for both immiscible and miscible flow. The grid is intentionally coarse for visualization purposes. (A) and (B) show immiscible fluids; (C) and (D) show miscible fluids.}
	\label{fig:2}
\end{figure}

To initialize the miscible simulations, we assume that the initial concentration field has an interface with a finite thickness of $1.5\Delta x$, where $\Delta x$ is the coarse grid resolution (see \S\ref{sec:nummethods} and figure~\ref{fig:2}). Hence, no discontinuous material contrasts arise in our miscible computations.

\subsubsection{Numerical Methods} \label{sec:nummethods}
We discretize the governing equations, \eqref{eq:GE1} and \eqref{eq:GE2}, using the numerical method derived, verified and validated in \citet{Qin2017}. Building on \citet{Suckale2010a}, and \citet{Suckale2012}, we developed this computational technique specifically for multiphase flow problems with large viscosity contrasts at low Reynolds number. It consists of three main components. First, a multiphase Navier-Stokes solver that handles substantial and discontinuous differences in the properties of material phases by adopting an implicit implementation of the viscous term, time-step splitting, and an approximate factorization of the sparse coefficient matrices for computational efficiency. Second, a level-set-based interface solver that tracks the motion of a fluid-fluid interface through an iterative topology-preserving projection \citep{Qin2015}. The sharp interface solver is only pertinent if the two fluids are immiscible. In the computationally simpler case of two miscible fluids, we solve an advection-diffusion equation for concentration. The third component is an adaptive grid refinement algorithm that increases resolution in the vicinity of the moving interface where most of the salient physical processes originate. Accurate interface advection is highly dependent on grid resolution \citep[e.g.,][]{sethian2003level}, particularly in flows that hinge sensitively on the growth of interface instabilities. To maximize numerical resolution at the interface, we adopt an adaptive mesh refinement strategy that tracks the interface position over time. Figure~\ref{fig:2} shows a close-up of the computational domain around the interface to schematically illustrate the grid refinement at the initial condition (figure~\ref{fig:2}A) and at a later time (figure~\ref{fig:2}), both shown at intentionally coarse resolution for easier visualization. The refined zone extends as the interface is stretched by flow. While the computational challenges associated with tracking diffusive interface are less pronounced, we adopt the same grid refinement strategy for the miscible case (figures~\ref{fig:2}C and D). For more details regarding the numerical technique, including the various benchmarks performed to verify and validate the numerical method, please refer to \citet{Qin2017} and the online supplement to this study.

%%%%%%%%%%%%%%%%%%%%%%%%%%%%%%%%%%%%

\begin{sidewaystable}
	\centering
	\caption{\small \textbf{Summary of analogue experimental data, and numerical and analytical results.} Viscosity ratio ($M$), density contrast ($\Delta\rho$), tube radius ($R$), tube length ($L$), front rise speed ($u$), Poiseuille number (Ps), dimensionless core radius ($\delta$) and flow regime (Reg) observed in experiments (EXP) and numerical models (NUM). Thin-core, $\delta_\mathrm{thin}$, and thick-core, $\delta_\mathrm{thick}$, radii, and flow reversal flux ratio ${\mathrm{Te_{rev}}}/\mathrm{Te}$ in analytical models (ANA). Bold numbers indicate the solution observed in corresponding experiments and simulations.} \label{tab:1} 
	\tabcolsep=5pt
	\begin{tabular}{llccccccccccccc}
		\hline
		\# & Analogue fluids & Reg & $M$ & $\Delta\rho$ & $R$ & $L$ & $u$\tiny $/10^{-5}$ & Ps & $u$\tiny$/10^{-5}$ & Ps & $\delta$ & $\delta_\mathrm{thin}$  &  $\delta_\mathrm{thick}$ &  ${\mathrm{Te_{rev}}}/\mathrm{Te}$ \\
		&&&&&&&\small(EXP)&\small(EXP)&\small(NUM)&\small(NUM)&\small(NUM) & \small(ANA)& \small(ANA)& \small(ANA) \\
		& & & - & kg m$^{-3}$ &  mm & mm &  m s$^{-1}$ & - &  m s$^{-1}$ &-&-&-&-&- \\
		\hline
		\#1 & Syrup / dil. syrup & \rom{2} &  68.46 & 77 & 7.2 & 720 & 2.51 & 0.065 & 1.16 & 0.030 & 0.55 & 0.26 & \textbf{0.61} & 0.10\\
		\#2 & Syrup / v.dil. syrup & \rom{1} & 18500 & 284 & 4 & 400 & 2.61 & 0.065 & 1.24 & 0.031 & 0.59 & 0.06 & \textbf{0.61} & 0.11\\
		\#3 & Dil. syrup / glycerol & \rom{3} & 1.27 & 125 & 4 & 400 & 22.0 & 0.023 & 11.48 & 0.012 & 0.52 & \textbf{ 0.53} & 0.60 & 0.01 \\
		\#4 & Dil. syrup / glycerol & \rom{3} & 2.34 & 147 & 4 & 400 & 17.1 & 0.032 & 8.02 & 0.015 & 0.52 & \textbf{0.52} & 0.60 & 0.04\\
		\#5 & Glycerol / water & \rom{1} & 1700 & 257 & 4 & 400 & 132 & 0.061 & 60.59 & 0.028 & 0.61 & 0.11 & \textbf{0.62} & 0.11\\
		\#6 & Syrup / dil. syrup & \rom{1} & 1371 & 144 & 4 & 400 & 2.27 & 0.067 & 0.91 & 0.027 & 0.59 & 0.12 & \textbf{0.61} & 0.11\\
		\#7 & Syrup / water & \rom{1} & 34636 & 423 & 7.2 & 720 & 36.8 & 0.065 & 17.55 & 0.031 & 0.59 & 0.05  & \textbf{0.61} & 0.11 \\
		\#8 & Syrup / water & \rom{1} & 30545 & 423 & 7.2 & 720 & 40.0 & 0.062 & 19.3 & 0.03 & 0.59 & 0.05 & \textbf{0.62} &0.11\\		
		\#9 & Syrup / glycerol & \rom{2} & 29.36 & 168 & 7.2 & 720 & 9.43 & 0.061 & 4.95 & 0.032 & 0.56 & 0.33 & \textbf{0.60} & 0.10\\
		\#10 & Dil Syrup / glycerol & \rom{3} & 2.11 & 123 & 7.2 & 720 & 99.5 & 0.036 & 46.99 & 0.017 & 0.52 & \textbf{0.44} & 0.68 & 0.10   \\
		\#11 & Dil. syrup / glycerol & \rom{3} & 2.11 & 123 & 4 & 400 & 27.3 & 0.032 & 13.65 & 0.016 & 0.52 & \textbf{0.51} & 0.61 & 0.04 \\
		\hline
	\end{tabular}
\end{sidewaystable}

\section{Results} \label{sec:results}

\begin{figure}
	\centering
	\includegraphics[width=0.7\textwidth]{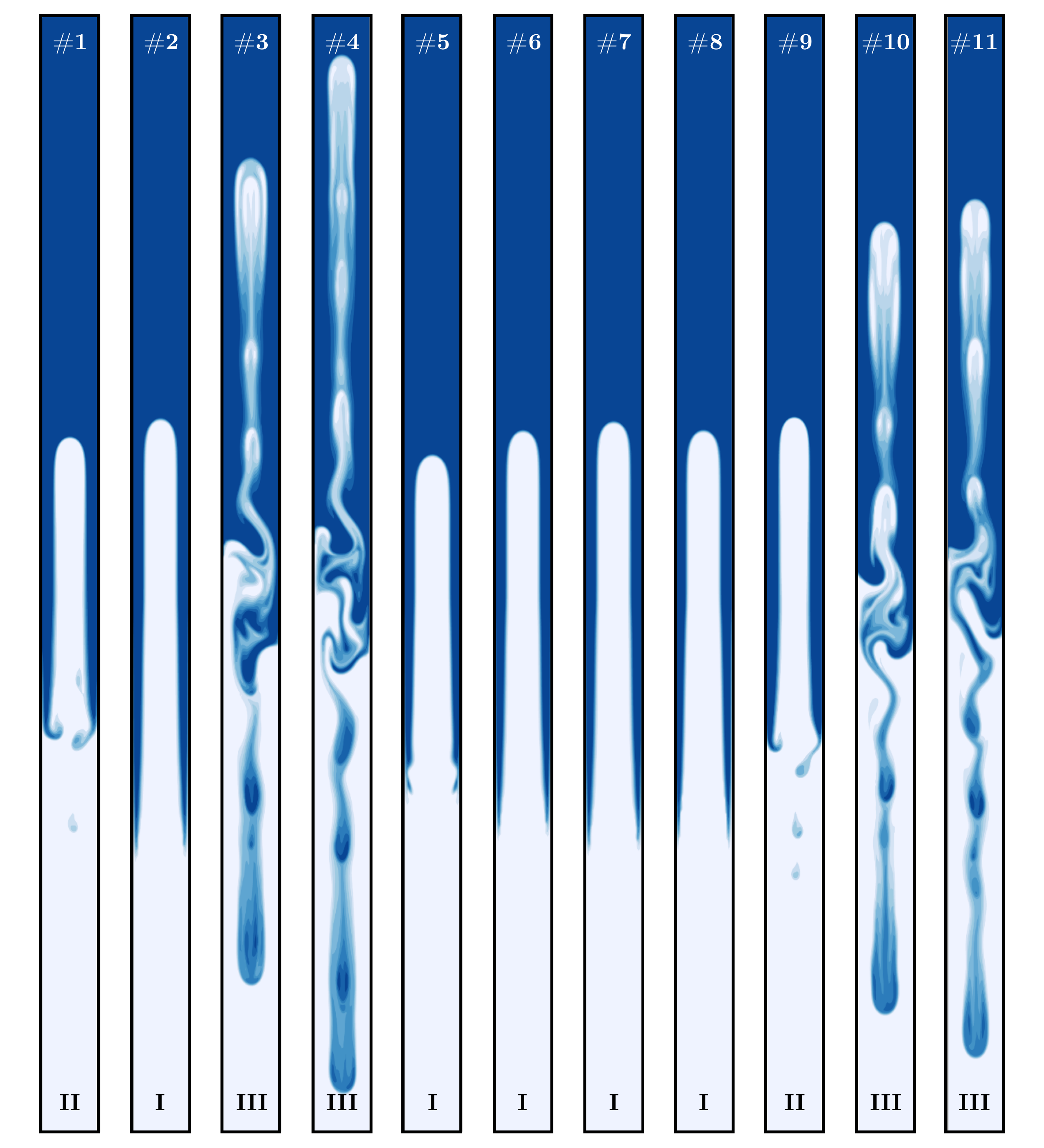}
	\caption{\small Numerical reproduction of all 11 experiments performed by \citet{Stevenson1998}. The aspect ratio of all laboratory tubes was 1:100 despite different physical lengths and widths. Here and after, only central part of numerical domain shown for better visibility.}
	\label{fig:3}
\end{figure}

\subsection{Virtual Reproductions of Analogue Experiments} \label{sec:results-stevenson}

\citet{Stevenson1998} initiate flow by inverting closed tubes filled with the denser, more viscous fluid (dark blue in our figures) in the lower half, and the buoyant, less viscous fluid (light blue in our figures) in the upper half. We initiate our models with the two fluids in an unstable density stratification and a slight cosine perturbation along their interface (see figure~\ref{fig:2}). We have verified that an initial condition of different symmetry yields qualitatively equivalent model outcomes, as shown in the online supplement. To represent the rigid glass walls of the experimental test tubes, we set all four side boundaries to no-slip ($\bf{v}=0$). All simulations are computed on a 40$\times$4000 grid with a factor 4 refinement applied to resolve the interface.

\begin{figure}
	\centering
	\includegraphics[width=0.75\textwidth]{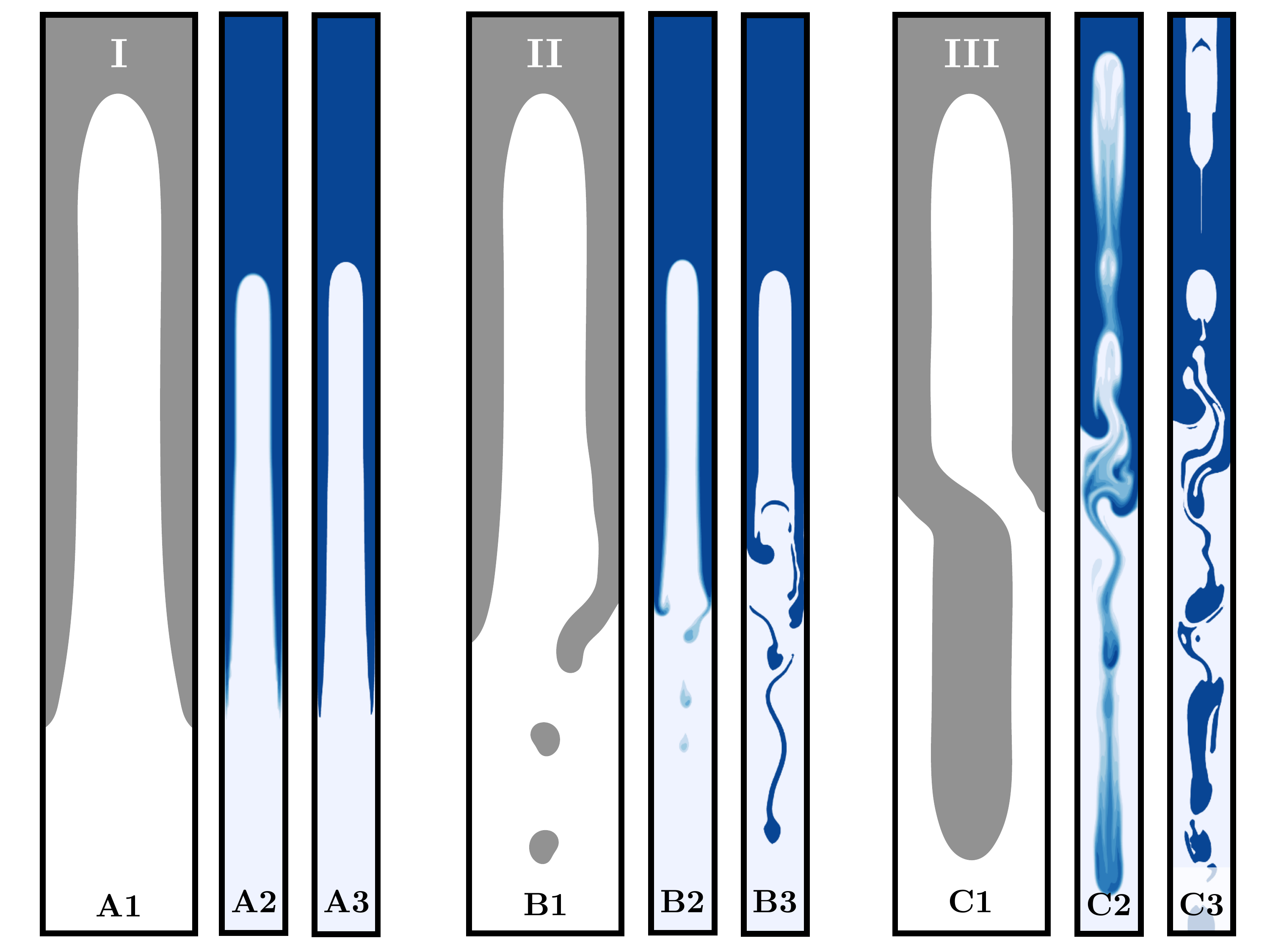}
	\caption{\small Direct numerical simulations of the three primary flow regimes (\rom{1}, \rom{2}, \rom{3}) observed in bidirectional tube flow for different viscosity contrasts. Miscible (A2, B2, C2) and immiscible (A3, B3, C3) flows shown in comparison to schematics (A1, B1, B2) reproduced from \citep{Stevenson1998}. Simulation snapshots given at $t = 200 \times t_0$ (flow regime \rom{1}: A2 and A3), $t = 200 \times t_0$ (flow regime \rom{2}: B2 and B3) and $t = 1000 \times t_0$ (flow regime \rom{3}: C2 and C3), respectively.}
	\label{fig:4}
\end{figure}
We numerically reproduce all eleven analogue experiments by \citet{Stevenson1998}, based on the detailed material properties they reported (see table \ref{tab:1}). Figure \ref{fig:3} shows the resulting flows at times, where the front rise speed of the overturn flow has reached an approximate steady state; these are $t = 200 \times t_0$ for experiments \#1,2 and \#5--9, and $t = 1000 \times t_0$ for \#3,4 and \#10,11, where $t_0=R/U$ is the dimensionless, viscous time scale. 

We generally observe the same behavior as reported by \citet{Stevenson1998}, which they classified into three different overturn styles. In figure~\ref{fig:4}, we illustrate these three overturn styles in the reproduced experiments \#8, \#9, and \#10. At high viscosity ratios (${\mathrm{M}} > 300$)
%JS: The symbols ~> or ~< do not render as approximately greater or smaller than, but rather introduce a forced space. Is the space what you intended? If not, the appropriate symbol codes need to be added.
, the flow configuration is characterized by stable core-annular flow (flow regime I; figures~\ref{fig:4}A1-3). With decreasing viscosity contrast, interface waves become more pronounced. However, at intermediate viscosity ratios ($10<{\mathrm{M}} < 300$) the amplitude of interface waves remains small enough to avoid wave bridging and disintegration of the core (flow regime II; figures~\ref{fig:4}B1-3). Rather, the descending fluid intermittently rips off the tube wall, while the ascending fluid forms a coherent core in the center of the tube. Once the viscosity contrast becomes small (${\mathrm{M}} < 10$), wave bridging occurs and, as a consequence, both fluids sink or rise as separate batches in the center of the tube (flow regime III; figures~\ref{fig:4}C1-3). We note that all three flow regimes exhibit interface waves, but their amplitude and dynamic significance decrease as the viscosity contrast increases in agreement with theory \citep{Hickox1971}. The transition between these three flow regimes is thus more gradual than this categorization into three distinct regimes may suggest.

Figure~\ref{fig:4} shows the results of miscible simulations (figures~\ref{fig:4}A2, B2, and C2) in comparison to three equivalent but immiscible cases (figures~\ref{fig:4}A3, B3, and C3). The three flow regimes appear qualitatively similar for both miscible and immiscible fluids. Since the thickness of the mixing boundary layer is small in comparison to the flow features, this finding is not particularly surprising. The miscible flows tends to have less variability along the interface, which confirms that even a small degree of miscibility notably dampens the growth rate of interface instabilities \citep[e.g.,][]{meiburg2004density}.

\begin{figure}
	\centering
	\includegraphics[width=0.75\textwidth]{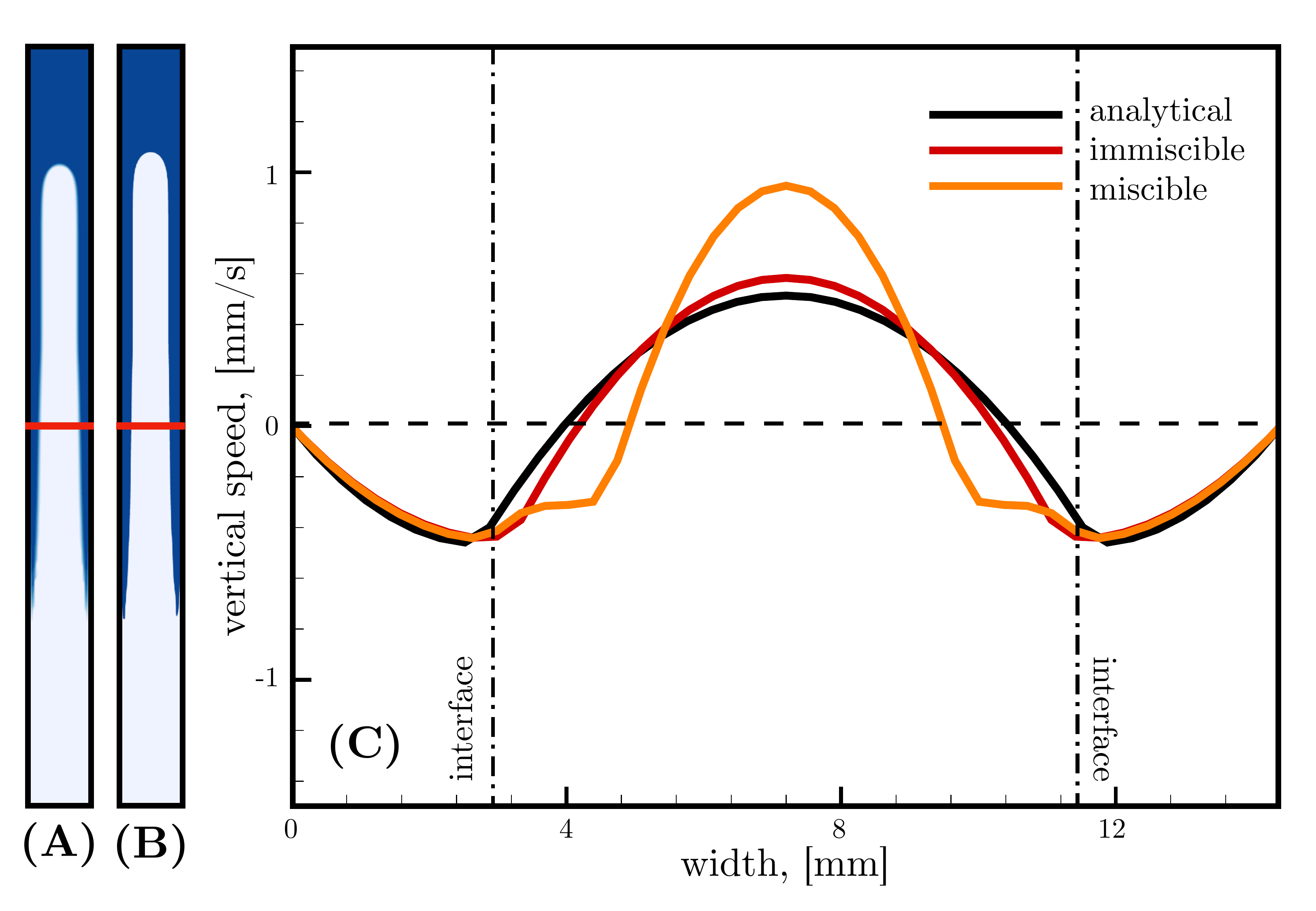}
	\caption{\small Reproduction of experiment \#8, treating the fluids as miscible (A) and immiscible (B). Numerical speed profiles (C) taken on marked cross sections (red bars in (A) and (B)) for miscible (yellow line) and immiscible (red line) flow compared to the analytical solution (black line).}
	\label{fig:5}
\end{figure} 

The availability of an analytical solution, at least in the limit of steady core-annular flow, provides a further opportunity to validate our numerical method beyond the benchmarks and tests presented in \citet{Qin2017}. In figure~\ref{fig:5}, we compare the virtual reproduction of experiment \#8 (figure~\ref{fig:5}A) for miscible fluids with its equivalent but immiscible case (figure~\ref{fig:5}B), and with the analytical solution calculated for the same flow parameters. To compare the numerical and analytical results, we take horizontal profiles of the vertical flow speed across the numerical domain in an area of well-developed core-annular flow. The flow profile of the immiscible simulation agrees remarkably well with the analytical result (figure~\ref{fig:5}C). The near fit between the immiscible numerical and the analytical solution is particularly encouraging considering that the latter was derived for steady state, while the former clearly arises in a transient flow. 

The vertical speed profile, however, is significantly modified by fluid miscibility (figure~\ref{fig:5}C). The two reasons are a gradual change in the density and viscosity across the mixing layer and a more distributed shear stress in the interfacial zone. Despite the small amount of mixing here, the dynamic consequences are notable both in the increased maximum rise speed in the core, as well as in the narrower upwelling portion of the flow. We find that an exponential variation of viscosity with composition further magnifies this effect (see online supplement).
% JS: Attempted to capture our discussion yesterday. Please double check and refine or adapt if needed.
%We thus note that fluid miscibility may be a highly consequential feature for examining the flow dynamics in natural systems.

\subsection{Analysis of Virtual Experiments} \label{sec:analysis-stevenson}
\begin{figure}
	\centering
	\includegraphics[width=0.75\textwidth]{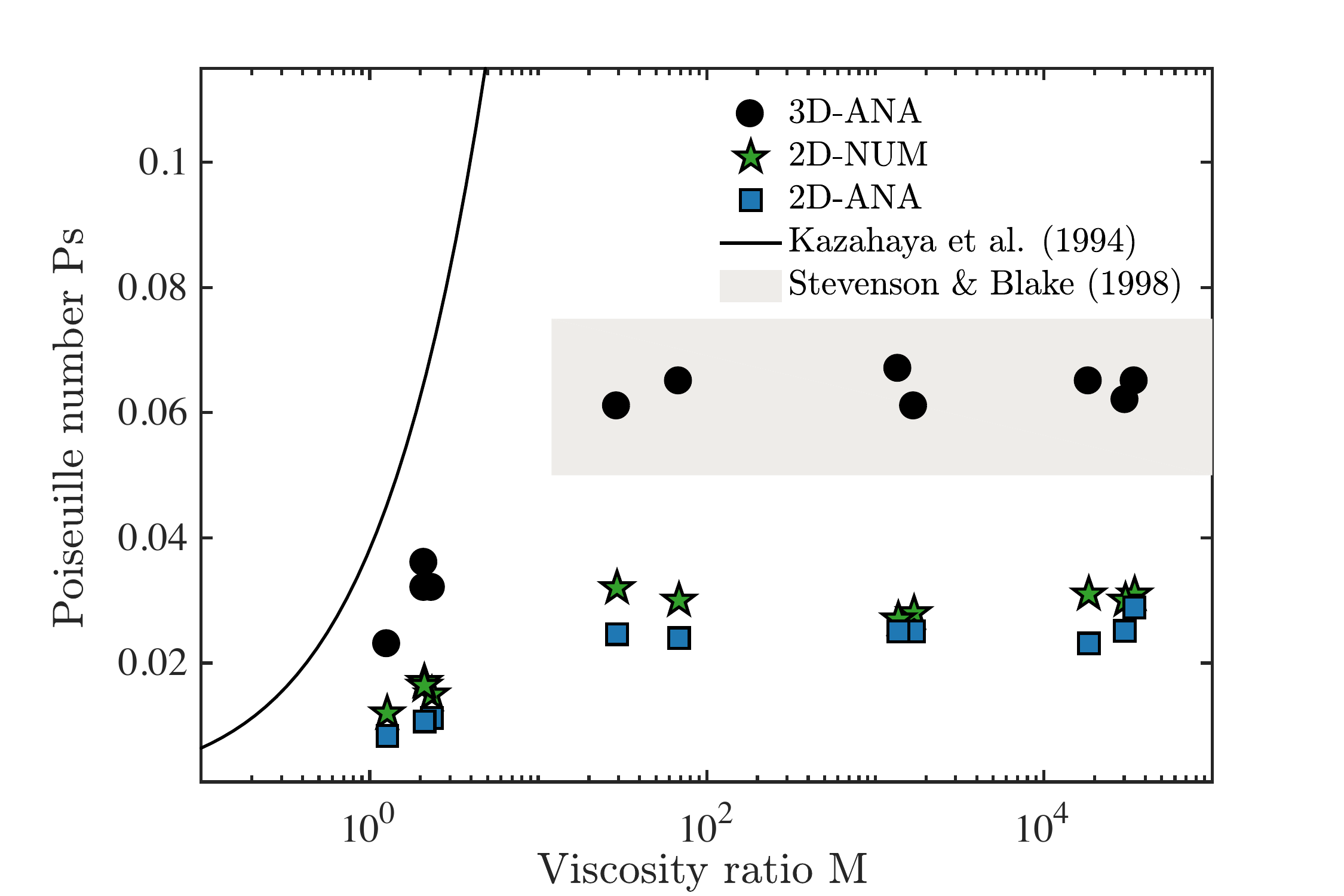}
	\caption{\small Plot of the Poiseuille Number, Ps, against the decadal logarithm of viscosity ratio M, for numerical simulations (green stars), analytical model predictions (black circles for 3D, blue squares for 2D), and the range of  analogue experiments (grey shading) by \citet{Stevenson1998}. Analytical solution by \citet{Kazahaya1994} shown for comparison (black line).}
	\label{fig:6}
\end{figure}

One of the main findings in \citet{Stevenson1998} is that the Poiseuille number in their experiments is essentially constant (i.e., Ps $\approx 0.065$) at finite viscosity contrast (approximately $\mathrm{M} > 10$). This behavior is surprising given the five orders of magnitude variation in viscosity contrast tested in the experiments. It also stands in stark contrast to the theory of \citet{Kazahaya1994}, which predicts a monotonic increase of Ps with M. Figure~\ref{fig:6} compares the range of experimentally determined Ps values with the numerical values resulting from our simulations, and the theoretical values calculated from our analytical model in 2D and 3D. We also show the theoretical values of \citet{Kazahaya1994} for comparison. Since our analytical model does not allow us to quantify the rise speed of a transient interface as done in \citet{Stevenson1998}, we instead compute Ps from the average vertical flow speed in the core phase. For consistency, we use the same measure to calculate Ps from the numerical results as well. The online supplement provides a comparison of these two metrics for rise speed and shows that they produce comparable values in our simulations.

Figure~\ref{fig:6} shows that our 3D analytical model (black dots) agrees well with the range of observed Ps numbers (grey shaded) reported by \citet{Stevenson1998} for $\mathrm{M} > 10$. It also quantifies the difference between Ps in 2D (blue squares) versus 3D (black circles). In 3D, the rise speed is about twice as fast as in 2D. The 2D analytical estimates (blue squares) agree well with the 2D numerical estimates (green stars). All three pieces, the measured data, the analytical values and the numerical outcomes, indicate that Ps levels off with increasing M, in contrast to \citet{Kazahaya1994}. 
\begin{figure}
	\centering
	\includegraphics[width=0.8\textwidth]{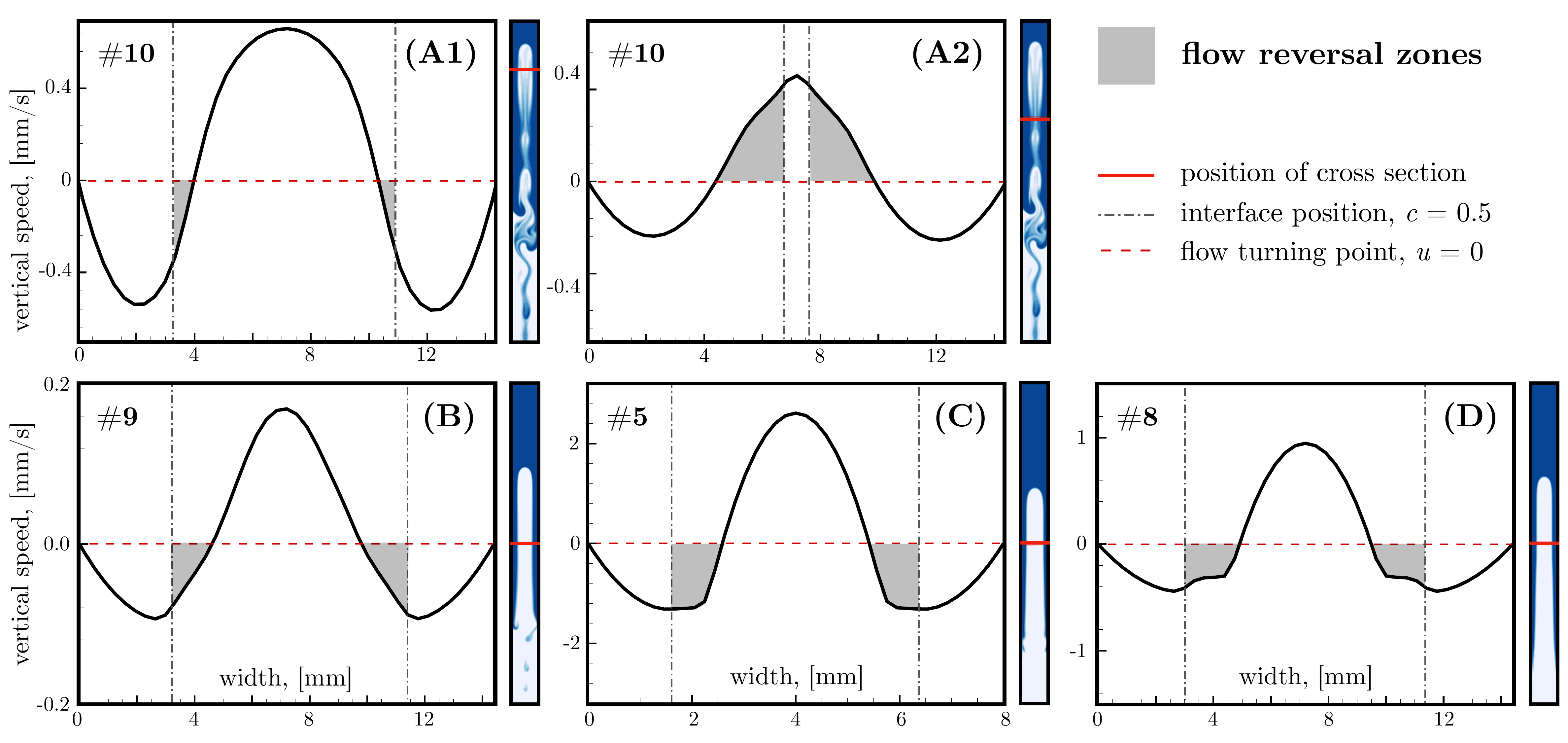}
	\caption{\small Zones of flow reversal (grey shaded) in bidirectional flow experiments on horizontal profiles of vertical speed, $u(r)$, along cross sections through virtual experiments marked by red bars across reported flow patterns. (A1), (A2): Profiles of experiment \#10, low viscosity contrast M, flow regime III. (B): Experiment \#9, intermediate M, regime II. (C): Experiment \#5, high M, regime I. (D): Experiment \#8, very high M, regime I.}
	\label{fig:7}
\end{figure} 

\citet{Stevenson1998} hypothesized that the disconnect between observed Ps values and model predictions by \citet{Kazahaya1994} is related to their assumption that the interface between the ascending and descending fluid is immobile (i.e., $u_i = 0$). Our virtual reproductions of the analogue experiments allow us to test this hypothesis. In figure~\ref{fig:7}, we plot vertical flow speed profiles across four different experiments: \#10 (figure~\ref{fig:7}A), \#9 (figure~\ref{fig:7}B), \#5 (figure~\ref{fig:7}C), and \#8 (figure~\ref{fig:7}D). These four experiments with increasing viscosity ratios represent the three different flow regimes indicated in figure~\ref{fig:4}. We arrange the experiments in this particular order to highlight the change in interface speed with increasing viscosity contrast. Our results give clear confirmation of the hypothesis by \citet{Stevenson1998}, showing that the interface is indeed not stationary.
%We have verified that this observation remains robust when considering immiscible fluids.

%The interface speed tends to be lowest at low viscosity contrast (i.e., experiment \#10 and \#9 in Fig.~\ref{fig:6}A1, A2 and B), increases with viscosity contrast (i.e., experiment \#5 in Fig.~\ref{fig:6}C), and tends to be highest at high viscosity contrast (i.e., experiment \#8 in Fig.~\ref{fig:6}D). 
A finite interface speed implies that the turning point between upward and downward oriented flow shifts into one of the two fluids. At low viscosity contrast (flow regime III), this shift may occur either in the ascending or in the descending fluid (see figures~\ref{fig:7}A1 and A2) and is linked to transient effects. For flow regimes I and II, the shift occurs in the ascending fluid in all simulations (figures~\ref{fig:7}B-D). This finding implies that, at intermediate to high viscosity contrasts (i.e., ${\mathrm{M}} > 10$), a portion of the buoyant fluid in fact flows downwards in the test tube---a phenomenon commonly referred to as flow reversal or backflow. %At small viscosity contrast (i.e., ${\mathrm{M}} <3$), flow reversal may happen in both the ascending (plot A$_1$) and the descending (plots A$_{2}$) fluid based on the position in the conduit. 

\begin{figure}
	\centering
	\includegraphics[width=0.8\textwidth]{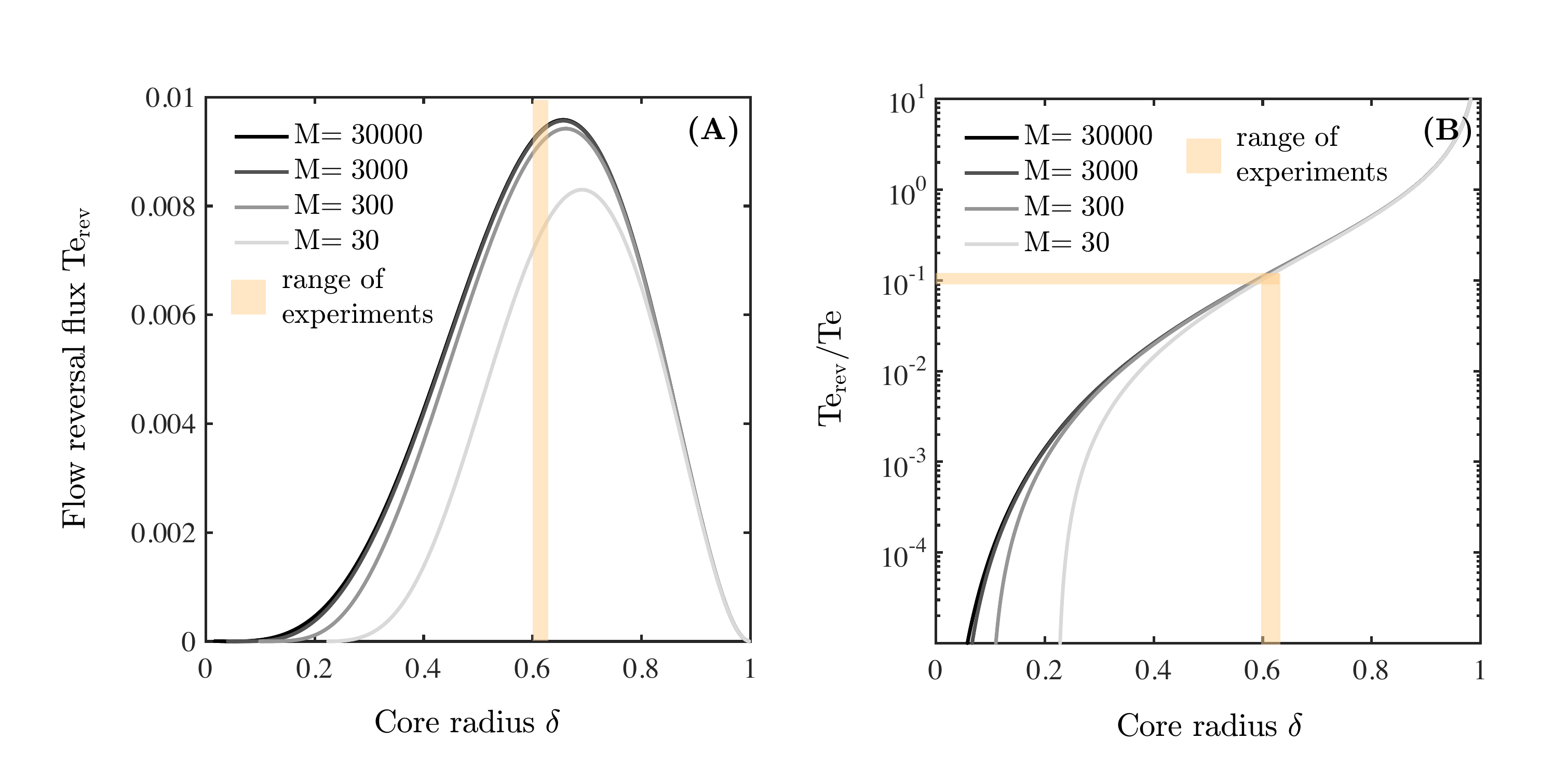}
	\caption{\small (B): Dimensionless flow-reversal flux in the ascending phase, ${\mathrm{Te_{rev}}}$, as a function of the core radius $\delta$. (A): Normalized dimensionless flow-reversal flux in the ascending phase, ${\mathrm{Te_{rev}}}$, scaled by the net dimensionless flux, Te, as a function of the core radius, $\delta$.}
	\label{fig:8}
\end{figure}

To quantify flow reversal more systematically, we compute the ratio between the flux in a given phase, Te, and the flow-reversal flux in the same phase ${\mathrm{Te_{rev}}}$, which we define as 
\begin{eqnarray}\label{eq:A11}
{\mathrm{Te_{rev}}}= \left|  \int_{{\delta_\mathrm{rev}}}^\delta 2\pi r u_a(r) \, dr  \right| \ .
% DP: I think, it should be the velocity of the ascending phase, u_a, that is integrated. It was previously written as u_d. Is my correction ok?
\end{eqnarray}
where $\delta_\mathrm{rev}$ is the point at which the vertical flow speed in the ascending fluid crosses zero, $u_a(\delta_\mathrm{rev})=0$. With this definition, ${\mathrm{Te_{rev}}}$ quantifies the flux of ascending fluid that is trapped in the flow reversal zone. Figure~\ref{fig:8} shows the analytically predicted flow reversal flux as a function of dimensionless model parameters. Figure~\ref{fig:8}A shows its dependence on the core radius, $\delta$, for different viscosity contrasts. For viscosity contrasts ${\mathrm{M}}> 10$, the experimentally observed core radii cluster just below the values resulting in maximum flow reversal. In figure~\ref{fig:8}B, we plot the ratio of flow reversal to total flux, again as a function of $\delta$ and ${\mathrm{M}}$. We find that ${\mathrm{Te_{rev}}}/$Te is mostly insensitive to the viscosity ratio M for $\delta>0.4$. Thus, we find that for all experiments in flow regimes I or II, the flow reversal flux should constitute about $10\%$ of the net flux.

The insight that the interface between the two fluids is inherently dynamic does not yet explain why Ps remains approximately constant with increasing M. To find an explanation, we return to \eqref{eq:A7} in \S\ref{sec:analyticmodel}. It states that Te is a quadratic function of ${\delta}$, which implies that two valid solutions for ${\delta}$ may exist for a given flux. The existence of multiple solutions is a common feature of multiphase flows, yet it does not necessarily imply that both solutions are realized in the laboratory \citep[e.g.,][]{Landman1991,Brauner1998,Ullmann2003,PicchiPoesio2016a}, or indeed in natural systems.

\begin{figure}
	\centering
	\includegraphics[width=0.9\textwidth]{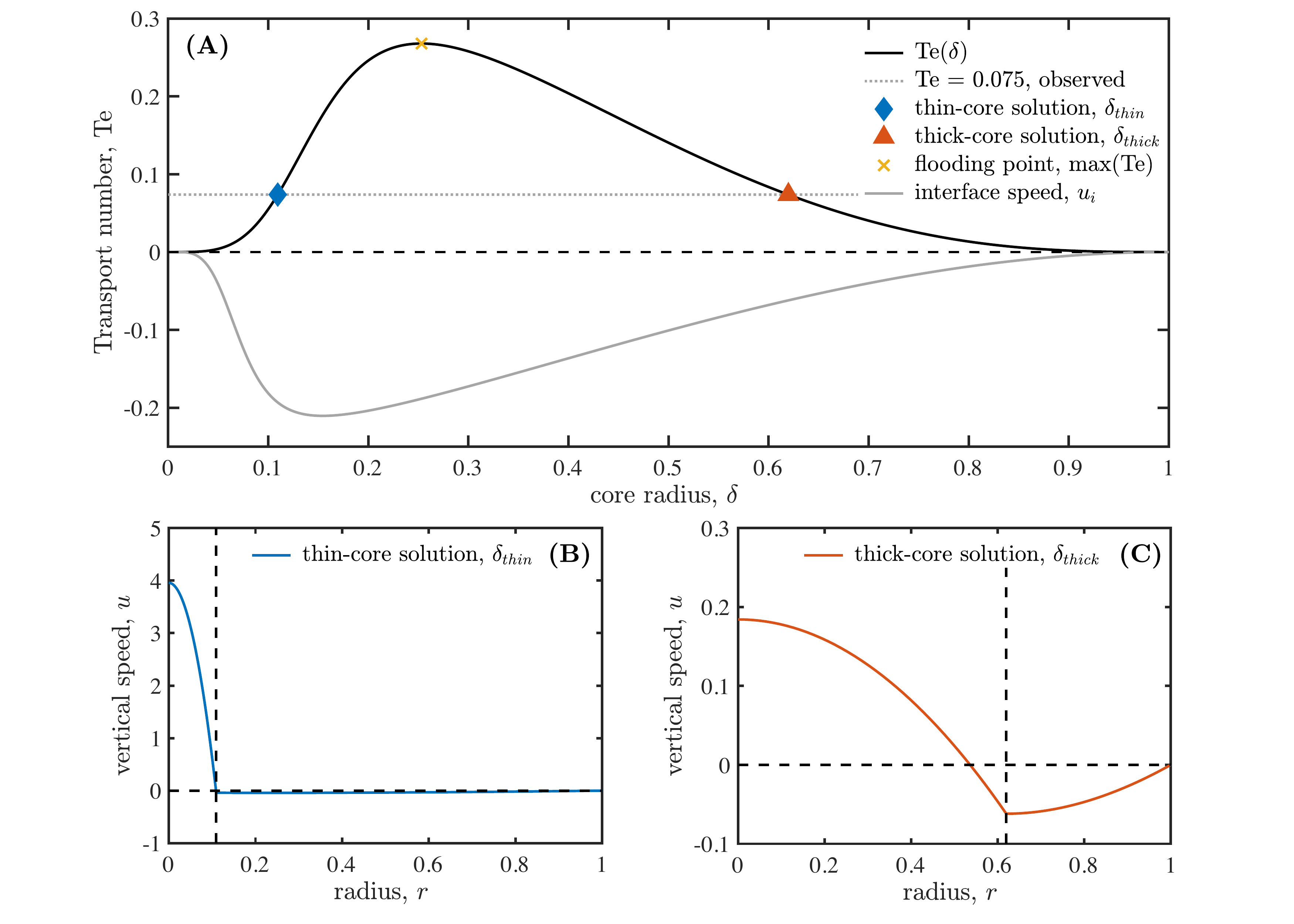}
	\caption{\small (A): Transport number, Te, and interface speed, $u_i$, as a function of the dimensionless core-radius, $\delta$, for experiment \#5 ($\mathrm{M}=1700$). (B) and (C): Dimensionless velocity profiles of the thin-core, $\delta_\mathrm{thin}$, and thick-core, $\delta_\mathrm{thick}$, solutions at $\mathrm{Te}=0.075$. We estimate Te based on the experimental rise speed.}
	\label{fig:9}
\end{figure}

In figure~\ref{fig:9}, we illustrate the two valid core-annular flow solutions for experiment \#5 (see table \ref{tab:1}). At the experimentally observed dimensionless flux, ${\mathrm{Te}} = 0.075$, our model predicts two solutions for the core radius, the thick-core ($\delta_\mathrm{thick}$, blue diamond) and the thin-core solution ($\delta_\mathrm{thin}$, red triangle). The two corresponding flow profiles, shown from the center of the tube to the wall in figures~\ref{fig:9}B and C, highlight that the same overall flux can be accommodated either by a thin, rapidly ascending core with a thick annulus, or through a thick, slowly ascending core with a thin annulus. Interestingly, both solutions are far removed from the point of maximum flux or flooding point (yellow $\times$). 
%At the flooding point, the kinematic wave speed switches sign as highlighted by the change in grey shading. For core radii larger than the flooding point, both the kinematic wave and the interface itself propagate downwards (i.e., $C_K<0$ and $u_i<0$). For core radii smaller than the flooding point, the kinematic wave and the interface typically propagate in different directions (i.e., $C_K>0$ and $u_i<0$).

%While two solutions are mathematically valid, only one can be realized in a stable flow field. 
Table \ref{tab:1} lists the two analytically computed core radii at the Te values inferred from the reported terminal rise speeds in all eleven experiments. The solutions compatible with their experimental and our numerical outcomes are printed in bold. We find that all experiments with viscosity contrasts of ${\mathrm{M}} > 10$ select the thick-core solution, $\delta_\mathrm{thick}$. The thin-core solution, $\delta_\mathrm{thin}$, may be pertinent for the experiments with comparable viscosities (experiments \#3, \#4, \#10 and \#11), but these cases do not exhibit stable core-annular flow and the applicability of the analytical model is thus questionable.

Figure~\ref{fig:10} shows analytical flux solutions for all experiments as a function of core radius and viscosity contrast. For each ${\mathrm{Te}}$--$\delta$ curve in figure~\ref{fig:10}A, we mark the solution that is realized in experiments ($\delta_\mathrm{thick}$, red triangle; $\delta_\mathrm{thin}$, blue diamond). We show curves for experiments \#3, \#4, \#10 and \#11 as light grey lines to convey that our analytical model is questionable for these cases. The experiments with ${\mathrm{M}} > 3$ (solid lines) all cluster around $\delta\approx 0.61$ and ${\mathrm{Te}}\approx 0.075$. The scatter in Ps number in \citet{Stevenson1998} (see figure~\ref{fig:6}) might hence not be entirely related to uncertainty of measurement, but reflect the fact that Te values at ${\mathrm{M}} > 3$ are predicted to be very similar but not identical (see figure~\ref{fig:10}B). For experiments with a viscosity ratio close to unity (light grey lines), Te assumes much lower values (figure~\ref{fig:10}B), and the two possible solutions are no longer clearly distinct.

This analysis suggests that the dichotomy in Ps number observed by \citet{Stevenson1998} and reproduced in figure~\ref{fig:6} reflects a shift from thick-core core-annular flow (flow regimes I and II) to unstable overturn flow (flow regime III). Contrary to the thin-core solution, the thick-core solution exhibits approximately constant core thickness over a large range of M (see figure~\ref{fig:10}A), which means that a constant Te translates to a constant Ps \eqref{eq:Te-Ps}. If any of the experiments were exhibiting the thin-core solution, Ps would increase with M.
%The dominance of the thick-core solution in the experiments by \citet{Stevenson1998} explains the observation that the Ps number is independent of viscosity contrast at finite viscosity contrast (i.e., M$>3$). 
%We find that the Ps number levels of with viscosity contrast, because all experiments at finite viscosity contrast (i.e., M$>3$) dynamically select the thick-core solution (see Fig.~\ref{fig:9}A). 

The result that only the thick-core solution, $\delta_\mathrm{thick}$, is realized at finite viscosity contrast in \citet{Stevenson1998} raises the question why this flow configuration appears to be preferable in this setting. This question is easiest to answer at high viscosity contrasts (i.e., ${\mathrm{M}} > 10,000$). As detailed in table~\ref{tab:1}, the predicted core radius for the thin-core solution, $\delta_\mathrm{thin}$, becomes as small as a few percent of the tube radius at very high viscosity contrasts (e.g., experiments \#2, \#7 and \#8). It is not surprising that a flow configuration with such a thin core is unstable. We recall that immiscible core-annular flow is always linearly unstable and only becomes metastable if the viscosity contrast is large enough to suppress interface waves to amplitudes that remain small compared to the core radius \citep[e.g.,][]{Hickox1971}. At very high M, the thin-core solutions are hence highly prone to wave bridging and flow collapse \citep[e.g.,][]{barnea1987unified}. 

\subsection{Simulations of Persistent Exchange Flow} \label{sec:results-general}
Determining the physical relevance of the two valid solutions based on core radius alone is unsatisfactory at intermediate viscosity contrast, when the two solutions, $\delta_\mathrm{thin}$ and $\delta_\mathrm{thick}$, become increasingly comparable. In this section, we explore the possibility that the physically pertinent solution may be controlled not only by the material parameters but also by the domain boundaries, an idea raised but not further explored by \citet{BarneaTaitel1985}. The experiments by \citet{Stevenson1998} were performed in closed test tubes, which is a significant difference to natural systems, where exchange flow is typically the consequence of continued flux.

To generalize insights obtained from closed to open systems, we perform simulations with open boundaries at the top and the base of the model domain. All of the simulations discussed in this section are forced by a time-independent inflow condition imposed along the base of the domain and set according to the analytical speed profile in 2D (see \eqref{eq:ud_2D}-\ref{eq:ua_2D}). We test both thin- and thick-core solutions by applying $u|_\mathrm{base}(r)=u_\mathrm{thin}(r)$ or $u|_\mathrm{base}(r)=u_\mathrm{thick}(r)$, respectively. We continue to enforce a no-slip condition ($\bf{v}|_\mathrm{side}=0$) along the side walls. For the outflow condition and the initial interface position we consider four different cases (see figure~\ref{fig:11}).

\begin{figure}
 	\centering
 	\includegraphics[width=0.8\textwidth]{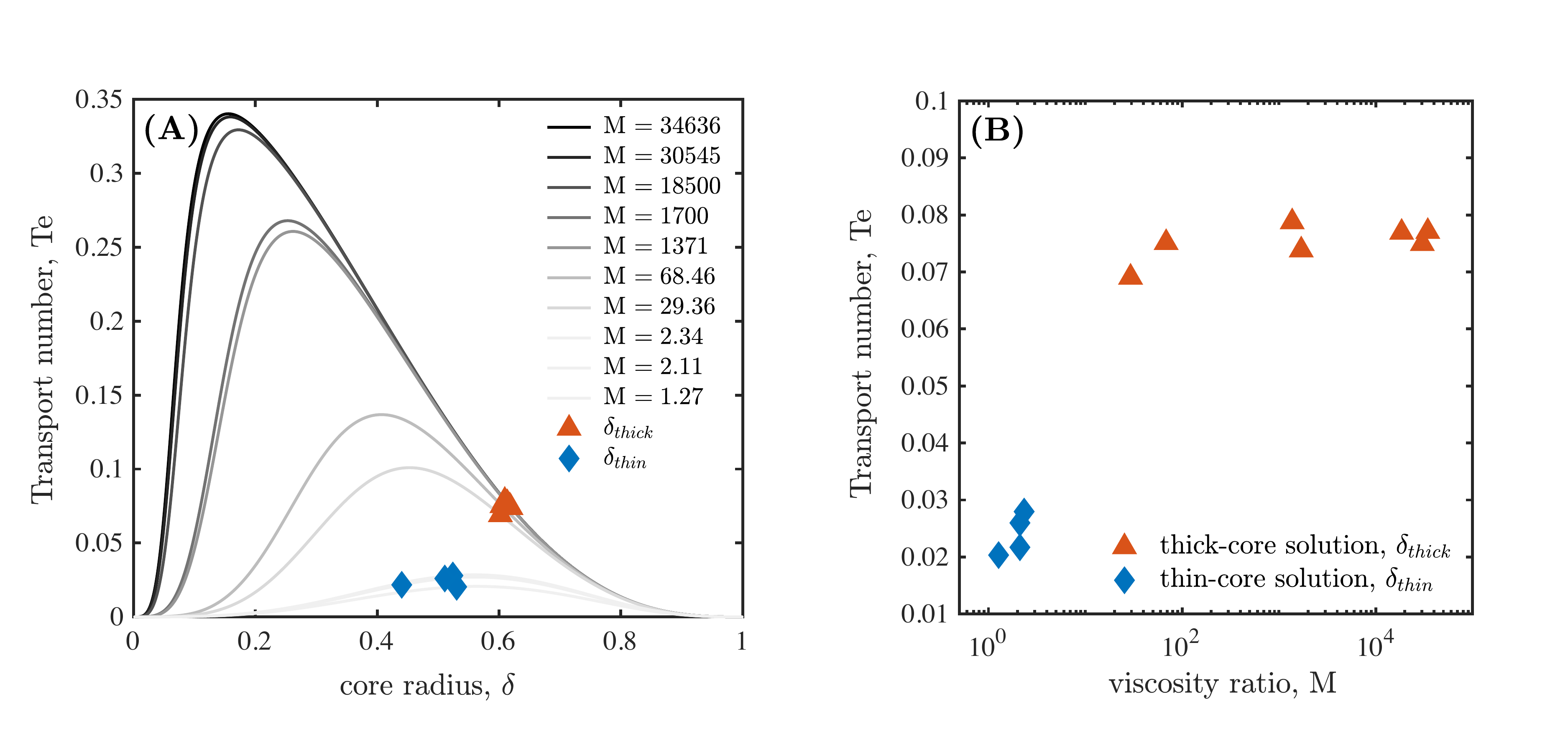}
 	\caption{\small (A): Te$-\delta$ curves for all experiments listed in table \ref{tab:1}. The plot shows ten instead of eleven curves, because experiments \#10 and \#11 have the same viscosity ratio. Numerically and experimentally realized core radii are highlighted as red triangles (thick-core) and blue diamonds (thin-core). (B): Transport number, Te, against the viscosity ratio, M, as computed from the analytical model for the experiments by \citet{Stevenson1998}.}
 	\label{fig:10}
\end{figure}

For the first case (figures~\ref{fig:11}A1-C), we impose a fixed outflow condition along the top boundary. We set the same analytical flow profile for the top and the base of the domain, i.e., $u|_\mathrm{base}(r)=u|_\mathrm{top}(r)=u_\mathrm{thick}(r)$, or $u|_\mathrm{base}(r)=u|_\mathrm{top}(r)=u_\mathrm{thin}(r)$, respectively. We also initiate the concentration field to the corresponding geometry extended through the whole domain (see figures~\ref{fig:11}A1 and B1). This choice implies that the interface is pinned both at the top and the bottom of the domain. While this setup is clearly contrived, it is an interesting end-member case for understanding the respective stability of the two solutions. Intuitively, one might expect that, if forced by the analytical solution on both ends, the flow field in the domain will remain close to that solution. Figure~\ref{fig:11}C shows that this is clearly not the case. While both interfaces are slightly wavy initially (figures~\ref{fig:11}A1 and B1), the thick-core solution stabilizes (figure~\ref{fig:11}A3). However, the thin-core case begins transitioning to the thick-core flow field from the top boundary downwards almost immediately after the onset of flow (figures~\ref{fig:11}B2 and B3). The speed profiles taken in the middle of the domain after simulations have reached an approximately steady state confirm that the flow field for both simulations---the one driven by $\delta_\mathrm{thick}$ and the one driven by $\delta_\mathrm{thin}$---closely approach the analytical thick-core solution (figure~\ref{fig:11}C).
 
\begin{figure}
	\centering
	\includegraphics[width=0.9\textwidth]{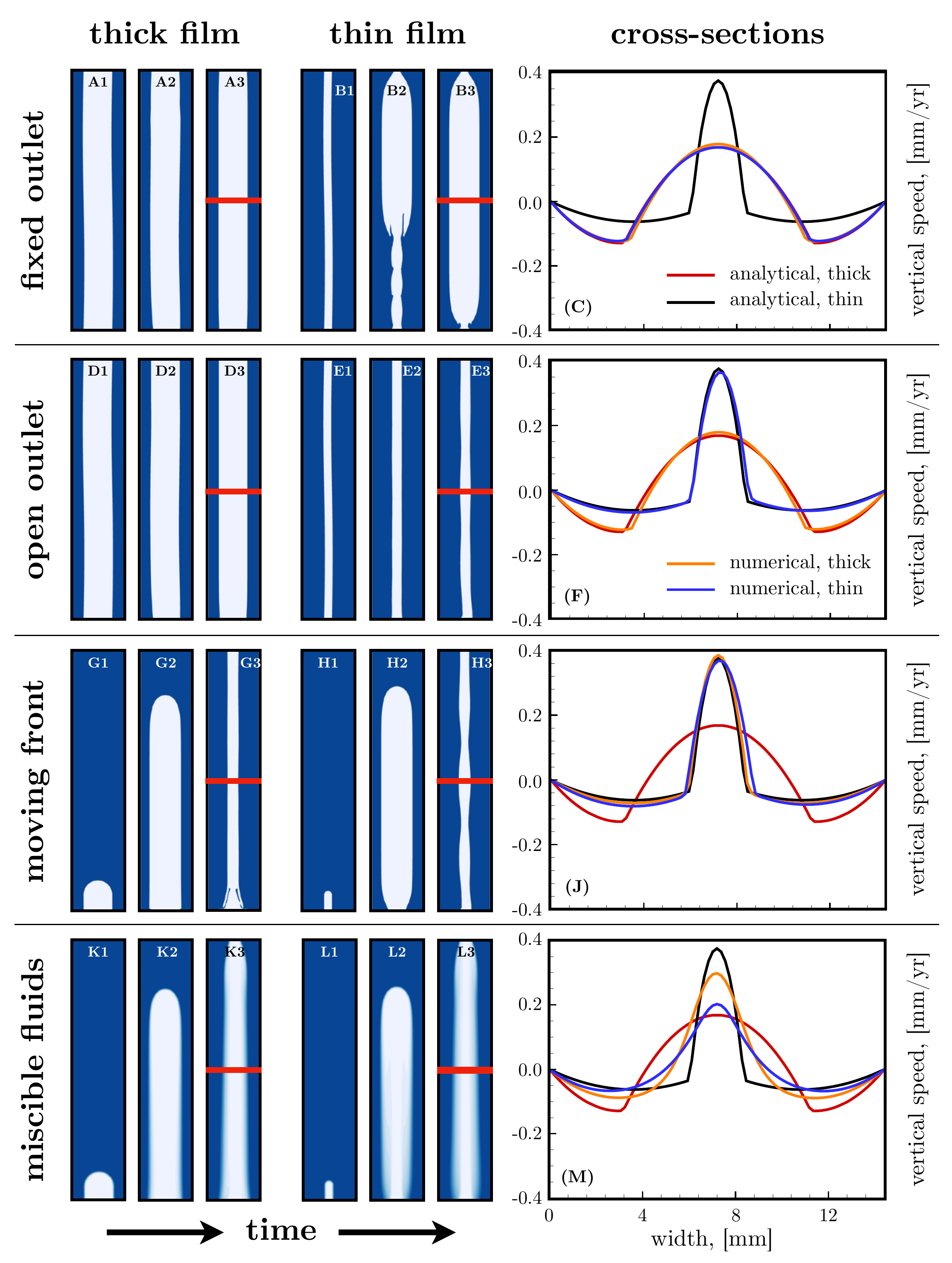}
	\caption{\small Snapshots of numerical simulations forced with thick- and thin-core analytical solutions at the inlet. Material properties are identical to experiment \#9 by \citet{Stevenson1998}. Results are shown for: fixed outlet boundary, enforcing the analytical model at the top boundary for thick- (A1--A3) and thin-core (B1--B3) solution; free outlet, stress-free top boundary allowing free flow through the top, starting from fully developed bidirectional configuration (D1--E3), and with a transient front moving through the domain (G1--H3); the same as previous case but with miscible fluids (K1--L3). Horizontal profiles of vertical speed (C, F, J, M) show numerical solutions approaching either thin- or thick-core analytical solutions depending on boundary conditions and fluid miscibility.}
	\label{fig:11}
\end{figure}

The second case (figures~\ref{fig:11}D1-F) represents a scenario with open outflow across a stress-free top boundary, which we enforce by setting $p|_\mathrm{top}=\mathrm{const.}$, and $\partial \mathbf{v}/\partial z |_\mathrm{top} = \mathbf{0}$. The flow field across the upper boundary is thus allowed to evolve over time, and the interface will move freely with the outflow. The initial position of the interface is identical to the first case (figures~\ref{fig:11}D1 and E1). The emerging flow fields correspond closely to the analytical solution imposed at the base of the domain. A thick-core inflow leads to stable thick-core configuration (figure~\ref{fig:11}D3) and a thin-core inflow leads to stable thin-core configuration (figure~\ref{fig:11}E3). This case demonstrates that the thin-core solution is a physically relevant flow configuration, at least at a viscosity contrast of ${\mathrm{M}} \approx 30$. Together with case one above, these simulations confirm that boundary conditions indeed have a profound influence on which mathematically valid flow configuration is realized in practice, as previously suggested in \citet{BarneaTaitel1985}.

One advantage of the variable over the fixed outflow case is that we can now consider a different initial interface position. In our third case (figures~\ref{fig:11}G1-J), we initiate an interface confined to the vicinity of the lower boundary (figures~\ref{fig:11}G1 and H1),  ensuring that the core radius corresponds to the analytical solution enforced at inflow. In these simulations, the interface moves through the domain in a similar way as in the experiments (figures~\ref{fig:11}G2 and H2), but eventually leaves the computational domain through the upper boundary. Immediately after onset, both the thick- and thin-core flows approach the thick-core solution (figure~\ref{fig:11}G2 and H2). However, as soon as the interface intersects with the upper boundary, the flow in both simulations collapses back to the thin-core solution (figures~\ref{fig:10}G3 and H3). This case demonstrates that the preferable flow configuration does not only depend on the boundary conditions, but that a transient effect on one boundary, such as the passage of the interface through the top, can trigger a switch in the flow configuration propagating across the entire domain.

Since the analytical solution strictly only applies to the immiscible limit, we have so far only considered immiscible fluids in this section. However, we have shown above (figure~\ref{fig:5}) that miscible flows are comparable to immiscible ones if the thickness of the mixing layer remains small compared to the flow features. In the fourth case (figures~\ref{fig:11}K1-M), we therefore repeat case three with miscible fluids. We observe a similar, but less pronounced behavior. Initially, the thin core thickens immediately (figure~\ref{fig:11}L2), but both flows collapse back to a thin-core solution once the interface passes the upper boundary (figures~\ref{fig:11}K3 and L3). The steady-state vertical speed profiles are reminiscent of the thin-core solution but more spread out (figure~\ref{fig:11}M). The interface also remains less wavy throughout, which supports the theoretical expectation that mixing along the interface stabilizes core-annular flow against interface waves \citep[e.g.,][]{meiburg2004density}. 
%The two speed profiles in figure~\ref{fig:11}M have diffused to different degrees for the two simulations, because the shown speed profiles refer to different physical times.
% JS: I double checked with Zhipeng that the speed profiles in (M) are in fact taken at the same time. It's not entirely clear to me why the thin-core case seems more well-mixed. I'm not sure we need to further comment on it though.

%------------------------------------------------

\section{Discussion} \label{sec:discuss}
\begin{figure}
	\centering
	\includegraphics[width=0.9\textwidth]{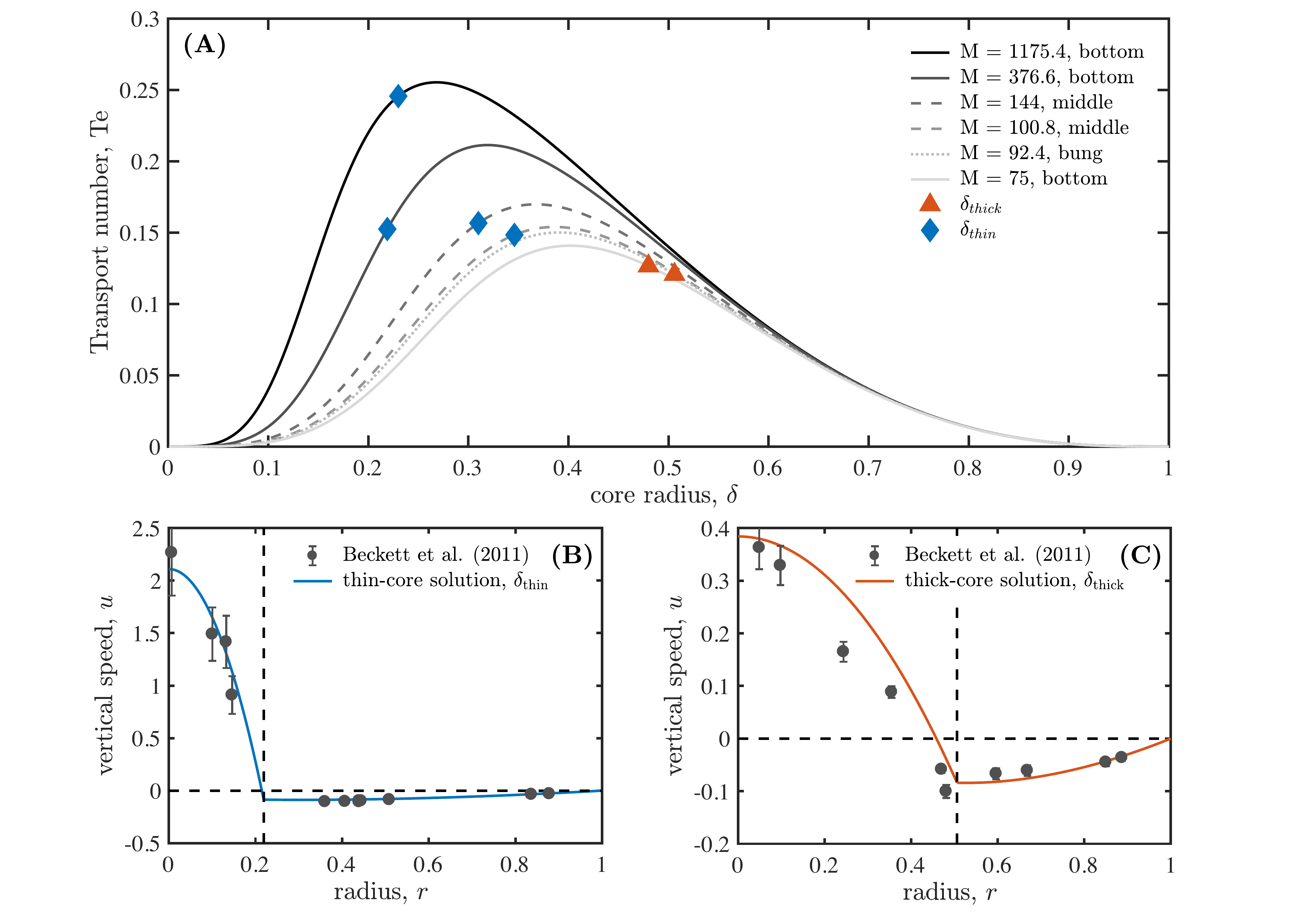}
	\caption{\small Reappraisal of laboratory experiments by \citet{Beckett2011}. (A): Te-$\delta$ curves and realized solutions for experiments \#9, \#11, \#15, \#16, \#17, \#20 based on material parameters provided in \citet{Beckett2011}. These xperiments were selected for exhibiting core-annular flow, with Frances Beckett kindly providing the measured data. (B): Fit of our analytical thin-film solution to the observational data for experiment \#15. (C): Fit of of our analytical thick-film solution to the observational data for experiment \#17.}
	\label{fig:12}
\end{figure}

\subsection{Theoretical ramifications}
\label{sec:modelprogress}
Laboratory experiments have provided invaluable insights into the dynamics of buoyancy-driven exchange flows, but they inevitably simplify the more complex flow problem they intend to represent. One aspect in which many analogue models of exchange flow differ fundamentally from natural or industrial systems is that bidirectional flow only occurs as transient behavior until a steady state of complete inversion of the two fluids is reached \citep{Stevenson1998,Huppert2007,Beckett2011,palma2011}. The steady state in the laboratory is hence very different from the steady state in many natural or industrial problems of interest. Our analysis suggests that this difference may be consequential, because closed systems only realize a subset of the possible flow solutions that occur in open systems. Experiments in closed domains may hence systematically underestimate the dynamic variability of open-system flow. 

We derive a simple analytical model that allows us to characterize the two steady-state solutions for core-annular flow. These two solutions refer to the same magnitude of flux, or Te, but differ in their core radii, $\delta_\mathrm{thick}$ and $\delta_\mathrm{thin}$, flow speed profile, and degree of flow reversal. We also predict the flux curve, $\mathrm{Te}(\delta)$, based only on the fluid properties and tube radius. The model does not require a fitting parameter to match experimental or numerical data, but does not predict the total flux nor the solution that is realized to achieve that flux. In fact, our analysis implies that it is not possible to predict these two outcome variables based on the fluid properties and the tube geometry alone, because the boundary conditions, above all pressure, play an important role in determining these.
%The experiments by \citet{Stevenson1998} suggest that buoyancy-driven exchange flows in closed tubes select the thick-core solution. The prominence of the thick-core solution in closed tubes explains the observation that the Ps number saturates with increasing viscosity contrasts (see figure~\ref{fig:6}). 

\citet{Huppert2007} suggested that buoyancy-driven exchange flow in vertical pipe tends to maximize flux, which would be equivalent to maximizing Te. This argument implies that the flux should at the flooding point for most or all of the experiments (figure \ref{fig:10}A), which is at odds with both the experimental and the numerical results discussed here. By inspection of table~\ref{tab:1} and figure \ref{fig:10}, we find that the flow system in this setup preferentially adjusts to the thick-core solution, despite the flux clearly not being maximized. Confirming \citet{Beckett2011}, we conclude that viscous dissipation is not useful to distinguish the two possible flow configurations. In fact, we find that the non-dimensional viscous dissipation is directly proportional to the flux magnitude Te, and thus depends on $\delta$ in the same fashion (see online supplement).

Our finding, that the boundaries and flow history select the realized solution, is consistent with laboratory experiments that mimic open-system, buoyancy-driven exchange flow by connecting a vertical tube to fluid reservoirs at both ends \citep[e.g.,][]{Huppert2007,Beckett2011,palma2011}. The flow patterns that arise in this more complex geometry are more varied than in \citet{Stevenson1998}. We have reanalyzed the measured velocity profiles shown in figure 9 from \citet{Beckett2011} and find that their experiments \#20, \#15 and \#11 (their figure 9 a,b, and c) correspond to the thin-core solution, experiment \#9 (their figure 9 d) is close to maximum flux where only one solution exists, and experiments \#17 and \#16 (their figure 9 e and f) realize the thick-core solution. Our analysis of their data is shown in figure~\ref{fig:12}, where we give the $\mathrm{Te}(\delta)$ curves and mark realized solutions for all experiments in their figure 9 (figure~\ref{fig:12}A). Our analytical model matches the measured speed profiles well for both the thin- and the thick-core example (figure~\ref{fig:12}B--C). The experimental data by \citet{Beckett2011} hence support our conclusion that both solutions are pertinent for understanding exchange flow in open systems.

\begin{figure}
	\centering
	\includegraphics[width=0.7\textwidth]{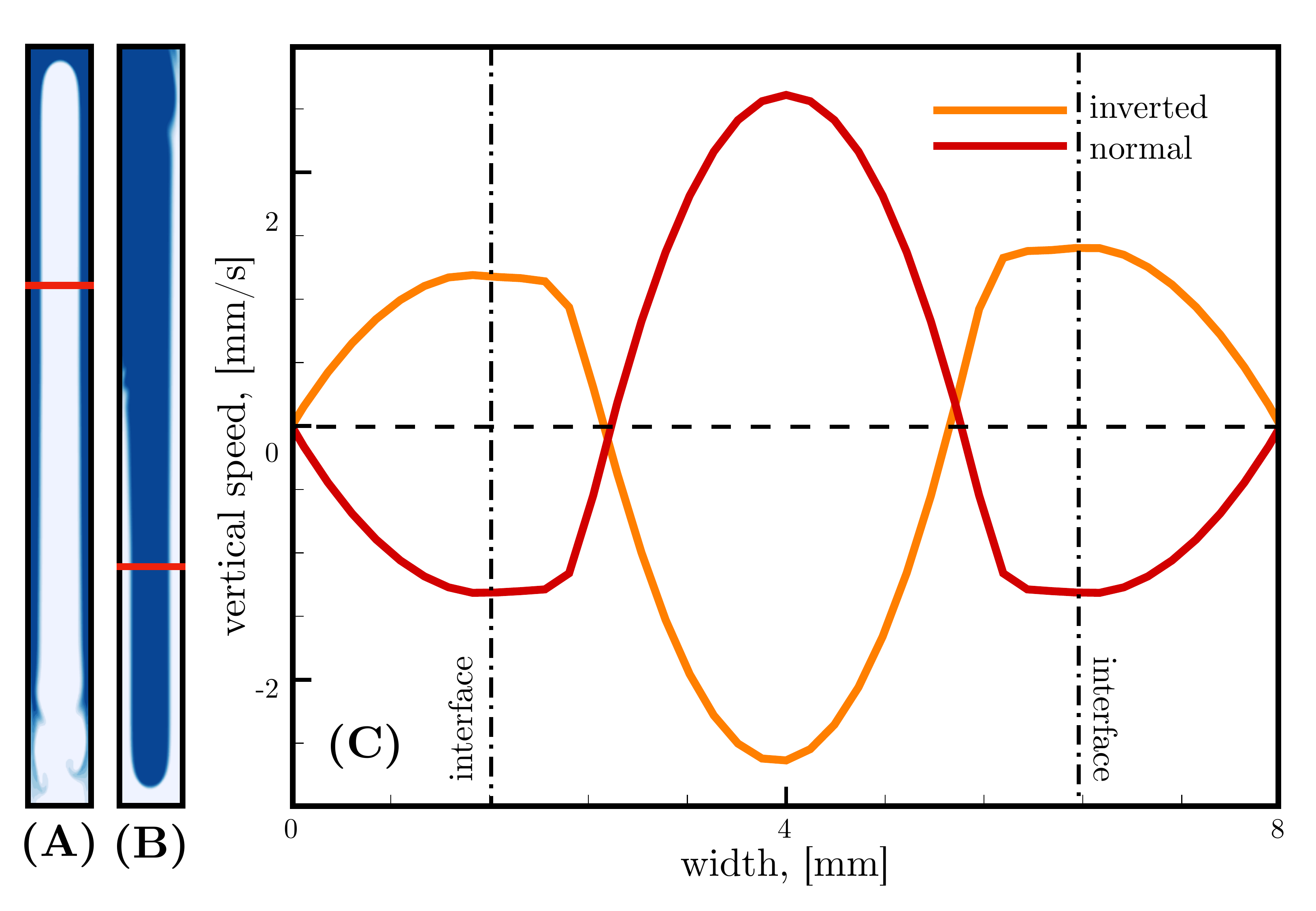}
	\caption{\small Comparison between the reproduction of experiment \#5 (A) and an additional simulation with inverted viscosity contrast $1/\mathrm{M}$ (B), where the heavy and now less viscous fluid (dark blue) forms a sinking core. (C): Horizontal  speed profiles for the inverted (yellow line) and normal (red line) viscosity contrast case.}
	\label{fig:13}
\end{figure}

\subsection{Ramifications for Volcanic Systems}
%\subsection{Towards a predictive, testable conduit flow model}
\label{sec:dataprogress}
In its current form, our model is not yet suitable for a direct quantitative comparison with field data from a specific volcano. Some of our insights, however, may inform the interpretation of field data on a qualitative level. One pertinent insight is that a change at either the base of a conduit or its opening in a volcanic crater could potentially trigger a switch in the flow regime throughout the conduit. The effect that different flow regimes in the volcanic conduit have on eruptive surface processes is well explored and reviewed by \citet{Vergniolle2000}. Our results suggest that the reverse is possible as well, i.e., eruptive surface processes can affect the flow regime in the volcanic conduit. For example, a disruption at the free surface, as might arise during an eruption or other events such as mass movements in the crater, could potentially trigger a switch in the flow configuration that is realized in the conduit. Of course, our simplistic boundary conditions (figures~\ref{fig:11}D1-E3 and G1-H3) do not adequately represent eruptive processes at a free surface. Nonetheless, our results demonstrate the potentially significant role of surface conditions in selecting a flow regime in the conduit. 
%One implication of this finding is that a change in the estimated or measured ascent speed of the magma does not necessarily require additional magma supply or a change in the volatile influx at depth. It could instead simply be related to a switch in the flow regime. 
%Moreover, as the expansion rate of compressible vapor bubbles is a function of the rise speed, a switch from thick- to thin-core regime may entail a further acceleration of the flow and thus possibly initiate a more eruptive episode. 
%JS: This connection in the last sentence above occured to me as I was editing this section. Not sure this reasoning is correct, please double check.
%TK: I'm leaning towards taking this out, simply because this would be the first time that we mention individual bubbles, the potential role of compressibility etc. I like the thought, but I think we would need to spend a bit more time evaluating this possibility.

We argue that the existence of two different, stable flow configurations could be reflected in erupted field samples from persistently degassing volcanoes, potentially from different stages of the same eruption. A switch from thick- to thin-core flow could increase the magma ascent rate by more than an order of magnitude, which may be detectable in microanalytical data. If a change in the estimated or measured ascent speed of the magma is detected, our results show that this observation does not necessarily require additional magma supply or a change in the volatile influx at depth. It could also be related to flow switching in the volcanic conduit. 

%We stress that the estimated vertical rise speeds in figures~\ref{fig:9}B and C were obtained for the laboratory setup and do not necessarily represent reasonable estimates for volcanic systems. The analytical model we derive (see \S\ref{sec:analyticmodel}), however, provides a versatile tool for estimating ascent speeds based on the physical parameters of actual volcanoes. 

Another potentially relevant insight of our study is the significant flow reversal present in all of the three flow regimes identified by \citet{Stevenson1998}. This result suggests that flow reversal might be the norm in volcanic conduits, rather than the exception. Our analytical model predicts that flow reversal occurs in the less viscous phase, a finding that is corroborated by our simulations (see figure~\ref{fig:13} and additional results in the online supplement). Typically, the buoyant, volatile-rich magma has the lower viscosity, as was assumed in the experiment by \citet{Stevenson1998}. In that case, flow reversal flux is oriented downward, which raises the question whether the magma trapped in flow reversal is simply cycled back into a crustal reservoir never to be sampled by eruption, or whether some magma may circulate in the conduit for some time before being erupted. In the latter scenario, continued cycling along with mixing between different magma batches may lead to fundamentally different compositional evolution of both the magma and its volatile load than would be expected from a straight decompression path.

%Potential avenues for sampling the flow-reversal zones include a switch in the flow regime, a disruption of the steady-state conduit flow as may arise in an eruption, or mixing along the interface between the two magmatic fluids.
% JS: Not sure about this sentence. The concept of "sampling the flow reversal zones" seems unrealistic. Perhaps we should rather express it in terms of "identifying the effects of flow-reversal"?

Flow reversal may also increase mixing between volatile-poor and volatile-rich magmatic melt, because the two fluids move in the same direction in some portion of the conduit. As demonstrated in figure~\ref{fig:5}B and C, even a small amount of mixing could have important dynamic ramifications for conduit flow.
%even a slight decrease of interface stresses from mixing translates into a potentially significant increase in the maximum rise speed. 
% JS: No need to repeat our debated argument here, I would say.
As pointed out by \citet{Witham2011}, magma mixing during conduit convection could be relevant for understanding why observed melt-inclusion trends from persistently degassing volcanoes rarely coincide with modeled degassing trends, as recently reviewed in \citet{metrich2008volatile}.

\section{Conclusions} \label{sec:conclude}
In this study, we reproduce, explain and generalize laboratory experiments of buoyancy-driven exchange flow in vertical tubes. We derive an analytical model for core-annular flow---the most commonly observed configuration of bidirectional flow---that is consistent with laboratory observations and direct numerical simulations. Our primary purpose in this paper is to understand the flow behavior observed in the laboratory, but our model may also provide a suitable starting point for integrating additional complexity needed for quantifying conduit flow in persistently degassing volcanoes. The key finding from our analysis is that core-annular flow is bistable at finite viscosity contrast. This result implies that buoyancy-driven exchange flow is not uniquely determined by the material properties of the fluids and geometric parameters of the tube. The pressure and fluxes at the boundaries of the domain, along with the transient history of the flow, play an important role in selecting the realized flow solution. 

\section*{Author Contributions}
Jenny Suckale conceptualized the study, integrated the results from the numerical and the analytical model, and wrote the paper. Zhipeng Qin performed the numerical simulations and composed the online supplementary. Davide performed the analytical derivation and computations. Tobias Keller assisted with the numerical modeling, creation of figures, and the text. Ilenia Battiato provided feedback on the study conception and the text.
%\begin{acknowledgments}

% JS: We shouldn't forget to include Acknowledgements for funding support etc.

\bibliographystyle{jfm}
\bibliography{./bi-direction}

\begin{thebibliography}{36}
\expandafter\ifx\csname natexlab\endcsname\relax\def\natexlab#1{#1}\fi
\def\au#1{#1} \def\ed#1{#1} \def\yr#1{#1}\def\at#1{#1}\def\jt#1{\textit{#1}}
  \def\bt#1{#1}\def\bvol#1{\textbf{#1}} \def\vol#1{#1} \def\pg#1{#1}
  \def\publ#1{#1}\def\arxiv#1{#1}\def\org#1{#1}\def\st#1{\textit{#1}}

\bibitem[Bai {\em et~al.\/}(1992)Bai, Chen \& Joseph]{bai1992lubricated}
{\sc \au{Bai, R.}, \au{Chen, K.} \& \au{Joseph, D.~D.}} \yr{1992}
  \at{Lubricated pipelining: stability of core�annular flow. part 5.
  experiments and comparison with theory}.  \jt{J. Fluid Mech}  \bvol{240},
  \pg{97--132}.

\bibitem[Barnea(1987)]{barnea1987unified}
{\sc \au{Barnea, D.}} \yr{1987}  \at{A unified model for predicting
  flow-pattern transitions for the whole range of pipe inclinations}.  \jt{Int.
  J. Multiph. Flow}  \bvol{13},  \pg{1--12}.

\bibitem[Barnea \& Taitel(1985)]{BarneaTaitel1985}
{\sc \au{Barnea, D.} \& \au{Taitel, Y.}} \yr{1985}  \at{Stability of annular
  flow}.  \jt{Int. Commun. Heat Mass}  \bvol{12},  \pg{611 -- 621}.

\bibitem[Beckett {\em et~al.\/}(2011)Beckett, Mader, Phillips, Rust \&
  Witham]{Beckett2011}
{\sc \au{Beckett, F.~M.}, \au{Mader, H.~M.}, \au{Phillips, J.~C.}, \au{Rust,
  A.~C.} \& \au{Witham, F.}} \yr{2011}  \at{{An experimental study of
  low-Reynolds-number exchange flow of two Newtonian fluids in a vertical
  pipe}}.  \jt{J. Fluid Mech}  \bvol{682}~(2011),  \pg{652--670}.

\bibitem[Brauner(1998)]{Brauner1998}
{\sc \au{Brauner, N.}} \yr{1998} {\em {Modelling and Experimentation in
  Two-Phase Flow}\/}, chap. {Liquid-Liquid two-phase flow},  \pg{pp. 221--279}.
   \publ{Springer Verlag}.

\bibitem[Burton {\em et~al.\/}(2007)Burton, Mader \& Polacci]{burton2007role}
{\sc \au{Burton, M.R.}, \au{Mader, H.M.} \& \au{Polacci, M.}} \yr{2007}
  \at{The role of gas percolation in quiescent degassing of persistently active
  basaltic volcanoes}.  \jt{Earth Planet. Sci. Lett}  \bvol{264}~(1),
  \pg{46--60}.

\bibitem[Dalziel(1992)]{dalziel1992maximal}
{\sc \au{Dalziel, S.~B.}} \yr{1992}  \at{Maximal exchange in channels with
  nonrectangular cross sections}.  \jt{J. Phys. Oceanogr.}  \bvol{22}~(10),
  \pg{1188--1206}.

\bibitem[Francis {\em et~al.\/}(1993)Francis, Oppenheimer \&
  Stevenson]{Francis1993}
{\sc \au{Francis, P.}, \au{Oppenheimer, C.} \& \au{Stevenson, D.}} \yr{1993}
  \at{{Endogenous growth of persistently active volcanoes}}.  \jt{Nature}
  \bvol{366}~(6455),  \pg{554--557}.

\bibitem[Frigaard \& Scherzer(1998)]{frigaard1998uniaxial}
{\sc \au{Frigaard, I.A.} \& \au{Scherzer, O.}} \yr{1998}  \at{Uniaxial exchange
  flows of two bingham fluids in a cylindrical duct}.  \jt{IMA. J. Appl. Math}
  \bvol{61}~(3),  \pg{237--266}.

\bibitem[Goyal \& Meiburg(2006)]{Goyal2006}
{\sc \au{Goyal, N.} \& \au{Meiburg, E.}} \yr{2006}  \at{{Miscible displacements
  in Hele-Shaw cells: two-dimensional base states and their linear stability}}.
   \jt{J. Fluid Mech}  \bvol{558},  \pg{329--355}.

\bibitem[Hickox(1971)]{Hickox1971}
{\sc \au{Hickox, C.~E.}} \yr{1971}  \at{{Instability due to Viscosity and
  Density Stratification in Axisymmetric Pipe Flow}}.  \jt{Phys. Fluids}
  \bvol{14}~(2),  \pg{251}.

\bibitem[Huppert \& Hallworth(2007)]{Huppert2007}
{\sc \au{Huppert, H.~E.} \& \au{Hallworth, M.~A.}} \yr{2007}
  \at{{Bi-directional flows in constrained systems}}.  \jt{J. Fluid Mech}
  \bvol{578},  \pg{95--112}.

\bibitem[Joseph {\em et~al.\/}(1997)Joseph, Bai, Chen \& Renardy]{Joseph1997}
{\sc \au{Joseph, D.~D.}, \au{Bai, R.}, \au{Chen, K.P.} \& \au{Renardy, Y.~Y.}}
  \yr{1997}  \at{Core-annular flows}.  \jt{Annu. Rev. Fluid Mech.}
  \bvol{29}~(1),  \pg{65--90}.

\bibitem[Kazahaya {\em et~al.\/}(1994)Kazahaya, Shinohara \&
  Saito]{Kazahaya1994}
{\sc \au{Kazahaya, Kohei}, \au{Shinohara, Hiroshi} \& \au{Saito, Genji}}
  \yr{1994}  \at{{Excessive degassing of Izu-Oshima volcano: magma convection
  in a conduit}}.  \jt{Bull. Volcanol}  \bvol{56}~(3),  \pg{207--216}.

\bibitem[Kuang {\em et~al.\/}(2003)Kuang, Maxworthy \& Petitjeans]{Kuang2003}
{\sc \au{Kuang, Jun}, \au{Maxworthy, Tony} \& \au{Petitjeans, Philippe}}
  \yr{2003}  \at{{Miscible displacements in cylindrical tubes: velocity fields
  and streamline patterns}}.  \jt{Eur. J. Mech. B Fluids}  \bvol{22}~(3),
  \pg{271--277}.

\bibitem[Landman(1991)]{Landman1991}
{\sc \au{Landman, M.J.}} \yr{1991}  \at{Non-unique holdup and pressure drop in
  two-phase stratified inclined pipe flow}.  \jt{International Journal of
  Multiphase Flow}  \bvol{17}~(3),  \pg{377 -- 394}.

\bibitem[Meiburg {\em et~al.\/}(2004)Meiburg, Vanaparthy, Payr \&
  Wilhelm]{meiburg2004density}
{\sc \au{Meiburg, Eckart}, \au{Vanaparthy, Surya~H}, \au{Payr, Matthias~D} \&
  \au{Wilhelm, Dirk}} \yr{2004}  \at{Density-driven instabilities of
  variable-viscosity miscible fluids in a capillary tube}.  \jt{Ann. N. Y.
  Acad. Sci.}  \bvol{1027}~(1),  \pg{383--402}.

\bibitem[M{\'e}trich \& Wallace(2008)]{metrich2008volatile}
{\sc \au{M{\'e}trich, Nicole} \& \au{Wallace, Paul~J}} \yr{2008}  \at{Volatile
  abundances in basaltic magmas and their degassing paths tracked by melt
  inclusions}.  \jt{Rev. Mineral. Geochem}  \bvol{69}~(1),  \pg{363--402}.

\bibitem[Palma {\em et~al.\/}(2011)Palma, Blake \& Calder]{palma2011}
{\sc \au{Palma, Jos{\'e}~L}, \au{Blake, Stephen} \& \au{Calder, Eliza~S}}
  \yr{2011}  \at{Constraints on the rates of degassing and convection in
  basaltic open vent volcanoes}.  \jt{Geochem. Geophys. Geosyst.}
  \bvol{12}~(11).

\bibitem[Petitjeans \& Maxworthy(1996)]{petitjeans1996miscible}
{\sc \au{Petitjeans, P} \& \au{Maxworthy, T}} \yr{1996}  \at{Miscible
  displacements in capillary tubes. part 1. experiments}.  \jt{J. Fluid Mech}
  \bvol{326},  \pg{37--56}.

\bibitem[Picchi \& Poesio(2016)]{PicchiPoesio2016a}
{\sc \au{Picchi, Davide} \& \au{Poesio, Pietro}} \yr{2016}  \at{Stability of
  multiple solutions in inclined gas/shear-thinning fluid stratified pipe
  flow}.  \jt{Int. J. Multiph. Flow}  \bvol{84},  \pg{176 -- 187}.

\bibitem[Qin {\em et~al.\/}(2015)Qin, Delaney, Riaz \& Balaras]{Qin2015}
{\sc \au{Qin, Zhipeng}, \au{Delaney, Keegan}, \au{Riaz, Amir} \& \au{Balaras,
  Elias}} \yr{2015}  \at{{Topology preserving advection of implicit interfaces
  on Cartesian grids}}.  \jt{J. Comput. Phys.}  \bvol{290},  \pg{219--238}.

\bibitem[Qin \& Suckale(2017)]{Qin2017}
{\sc \au{Qin, Z.} \& \au{Suckale, J.}} \yr{2017}  \at{{Direct numerical
  simulations of gas solid liquid interactions in dilute fluids}}.  \jt{{Int.
  J. Multiph. Flow}}  \bvol{96},  \pg{34--47}.

\bibitem[Ray {\em et~al.\/}(2007)Ray, Bunton \& Pojman]{Ray2007}
{\sc \au{Ray, E.}, \au{Bunton, P.} \& \au{Pojman, J.~a.}} \yr{2007}
  \at{{Determination of the diffusion coefficient between corn syrup and
  distilled water using a digital camera}}.  \jt{Am. J. Phys}
  \bvol{75}~(2007),  \pg{903}.

\bibitem[Russell \& Charles(1959)]{russell1959effect}
{\sc \au{Russell, TWF} \& \au{Charles, ME}} \yr{1959}  \at{The effect of the
  less viscous liquid in the laminar flow of two immiscible liquids}.  \jt{Can.
  J. Chem. Eng.}  \bvol{37}~(1),  \pg{18--24}.

\bibitem[Scoffoni {\em et~al.\/}(2001)Scoffoni, Lajeunesse \&
  Homsy]{Scoffoni2001}
{\sc \au{Scoffoni, J.}, \au{Lajeunesse, E.} \& \au{Homsy, G.~M.}} \yr{2001}
  \at{{Interface instabilities during displacements of two miscible fluids in a
  vertical pipe}}.  \jt{Phys. Fluids}  \bvol{13}~(3),  \pg{553--554}.

\bibitem[Sethian(1996)]{sethian2003level}
{\sc \au{Sethian, James~A}} \yr{1996} {\em Level set methods and fast marching
  methods\/}.  \publ{Cambridge University Press}.

\bibitem[Stevenson \& Blake(1998)]{Stevenson1998}
{\sc \au{Stevenson, David~S} \& \au{Blake, Stephen}} \yr{1998}  \at{Modelling
  the dynamics and thermodynamics of volcanic degassing}.  \jt{Bull. Volcanol.}
   \bvol{60}~(4),  \pg{307--317}.

\bibitem[Suckale {\em et~al.\/}(2010{\natexlab{{\em a\/}}})Suckale, Hager,
  Elkins-Tanton \& Nave]{Suckale2010}
{\sc \au{Suckale, Jenny}, \au{Hager, Bradford~H}, \au{Elkins-Tanton, Linda~T}
  \& \au{Nave, Jean-Christophe}} \yr{2010{\natexlab{{\em a\/}}}}  \at{{It takes
  three to tango: 2. Bubble dynamics in basaltic volcanoes and ramifications
  for modeling normal Strombolian activity}}.  \jt{J. Geophys. Res.}
  \bvol{115}~(B7),  \pg{B07410}.

\bibitem[Suckale {\em et~al.\/}(2010{\natexlab{{\em b\/}}})Suckale, Nave \&
  Hager]{Suckale2010a}
{\sc \au{Suckale, Jenny}, \au{Nave, Jean-Christophe} \& \au{Hager, Bradford~H}}
  \yr{2010{\natexlab{{\em b\/}}}}  \at{{It takes three to tango: 1. Simulating
  buoyancy-driven flow in the presence of large viscosity contrasts}}.  \jt{J.
  Geophys. Res.}  \bvol{115}~(B7),  \pg{B07409}.

\bibitem[Suckale {\em et~al.\/}(2012)Suckale, Sethian, Yu \&
  Elkins-Tanton]{Suckale2012}
{\sc \au{Suckale, Jenny}, \au{Sethian, James~a.}, \au{Yu, Jiun-der} \&
  \au{Elkins-Tanton, Linda~T.}} \yr{2012}  \at{{Crystals stirred up: 1. Direct
  numerical simulations of crystal settling in nondilute magmatic
  suspensions}}.  \jt{J. Geophys. Res.}  \bvol{117}~(E8),  \pg{E08004}.

\bibitem[Tan \& Homsy(1986)]{Tan1986}
{\sc \au{Tan, C~T} \& \au{Homsy, G~M}} \yr{1986}  \at{{Stability of miscible
  displacements in porous media : Rectilinear flow}}.  \jt{Phys. Fluids}
  \bvol{29},  \pg{3549--3556}.

\bibitem[Ullmann \& Brauner(2004)]{Ullmann2004}
{\sc \au{Ullmann, Amos} \& \au{Brauner, Neima}} \yr{2004}  \at{{Closure
  Relations for The Shear Stress in Two-Fluid Models for Core-Annular Flow}}.
  \jt{Multiphas. Sci. Tech.}  \bvol{16}~(4),  \pg{355--387}.

\bibitem[Ullmann {\em et~al.\/}(2003)Ullmann, Zamir, Ludmer \&
  Brauner]{Ullmann2003}
{\sc \au{Ullmann, A.}, \au{Zamir, M.}, \au{Ludmer, Z.} \& \au{Brauner, N.}}
  \yr{2003}  \at{Stratified laminar countercurrent flow of two liquid phases in
  inclined tubes}.  \jt{International Journal of Multiphase Flow}
  \bvol{29}~(10),  \pg{1583 -- 1604}.

\bibitem[Vergniolle \& Mangan(2000)]{Vergniolle2000}
{\sc \au{Vergniolle, S} \& \au{Mangan, M}} \yr{2000}  \at{{Hawaiian and
  Strombolian eruptions}}.  \bt{In {\em Encyclopedia of Volcanoes\/}},  \pg{pp.
  447--461}.

\bibitem[Witham(2011)]{Witham2011}
{\sc \au{Witham, Fred}} \yr{2011}  \at{{Conduit convection, magma mixing, and
  melt inclusion trends at persistently degassing volcanoes}}.  \jt{Earth
  Planet. Sci. Lett}  \bvol{301}~(1-2),  \pg{345--352}.

\end{thebibliography}

\newpage
\renewcommand{\thesection}{S} \setcounter{section}{0}
\renewcommand{\thetable}{S\arabic{table}} \setcounter{table}{0}
\renewcommand{\thefigure}{S\arabic{figure}} \setcounter{figure}{0}

\section{Additional discussion of numerical model}
	
This appendix provides: (1) An investigation on the impact of the initial condition on the bidirectional flow regimes; (2) A convergence test to assess the spatial and temporal resolution of the numerical model and a brief review of our adaptive grid refinement strategy; (3) A comparison of the average rise velocity used in the analytical model to the frontal rise speed used in analogue experiments; (4) A comparison of the linear dependence of viscosity on concentration to the nonlinear one; (5) The expression for calculating the viscous dissipation for concentric core-annular flows.

\subsection{Impact of the numerical initial condition on the bidirectional flow}
\begin{figure}[htb]
	\centering
	\includegraphics[width=0.8\textwidth]{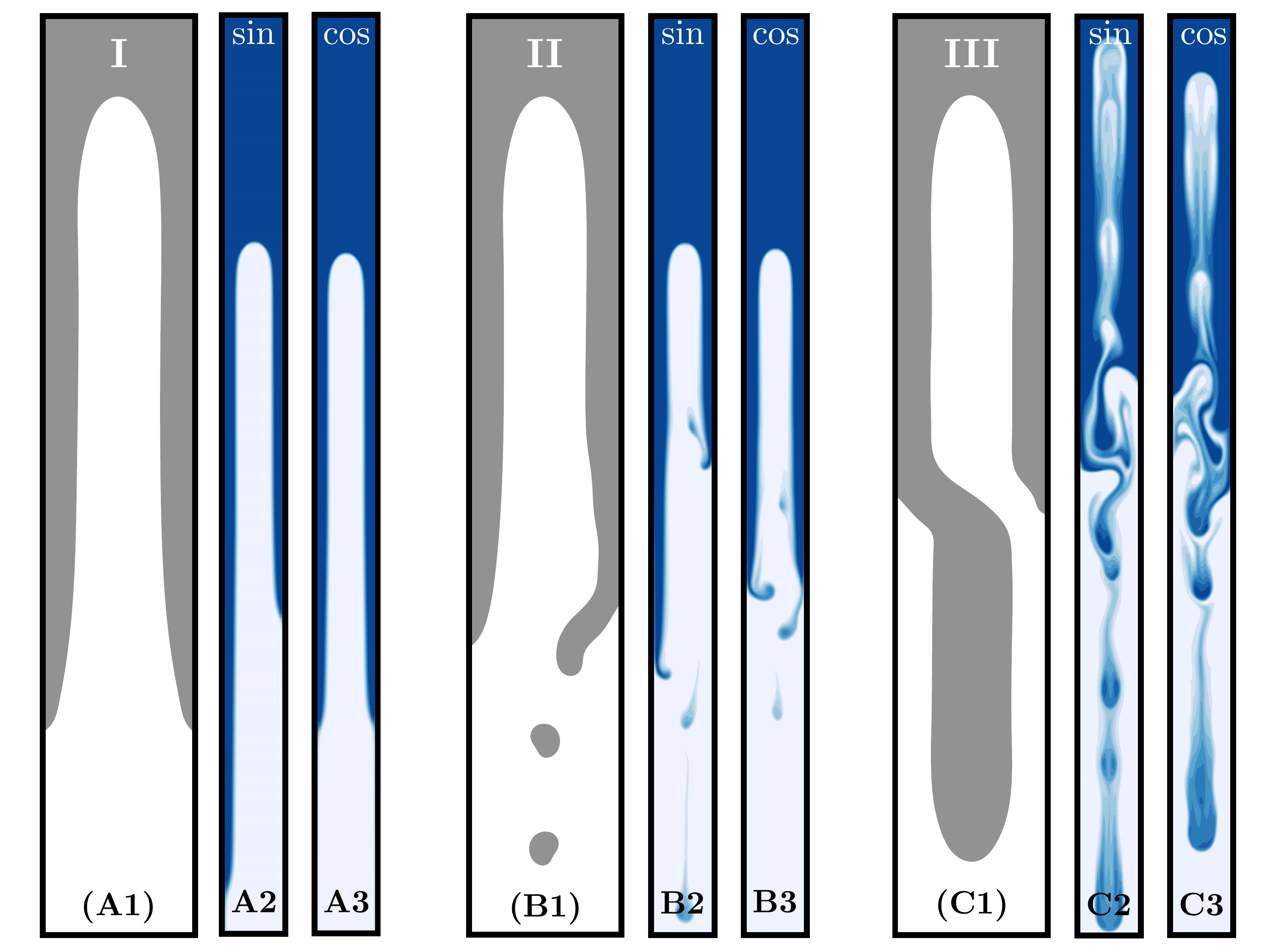}
	\caption{\small Direct numerical simulations of the three primary flow regimes observed in bidirectional pipe flow for different viscosity contrast with sine initial perturbation (A2, B2, C2) and cosine initial perturbation (A3, B3, C3)  in comparison to sketches (A1, B1, C1) reproduced from the analogue experiments by \citet{Stevenson1998}. The simulation snapshots are taken at 100s (flow regime \rom{1}: A2 and A3), 300s (flow regime \rom{2}: B2 and B3) and 200s (flow regime \rom{3}: C2 and C3), respectively. }
	\label{fig:S1}
\end{figure}

We reproduce the analogue experiments with two kinds of perturbations applied to the interface between the two liquids, a sine and a cosine function. The flow regimes with three different viscosity contrast are shown in figure \ref{fig:S1}. At high viscosity contrast (experiment \#8), steady core-annular flow forms from both initial perturbations. The descending annulus phase, however, is asymmetric when using a sine function initially (A2) in the sense that one side stretches further down than the other, whereas a cosine initial condition leads to a symmetric flow (A3). The degree of asymmetry depends on the magnitude of the initial perturbation. At intermediate viscosity contrast (experiment \#9), the simulations with both sine (plot B2) and cosine (plot B3) initial perturbations produce a similar flow regime where the ascending liquid rises in the center of the tube and the descending liquid breaks into small blobs. At low viscosity contrast (experiment \#10), the ascending and descending liquid is approximately symmetric for both initial conditions (figures \ref{fig:S1}C2-3). According to these results, the shape of the initial perturbation does not determine the flow regime, and will only have a minor effect on the flow symmetry. We also find that the simulations result in similar Ps number independent of their initial perturbation, even though the sine perturbation somewhat shortens the incubation time (see figure \ref{fig:S1}). Due to the negligible effect of the initial perturbation on the model outcomes, we only present the numerical simulations with a cosine initial condition in the manuscript.

\subsection{Numerical convergence test and adaptive grid refinement}  

\begin{figure}[htb]
	\centering
	\includegraphics[width=0.8\textwidth]{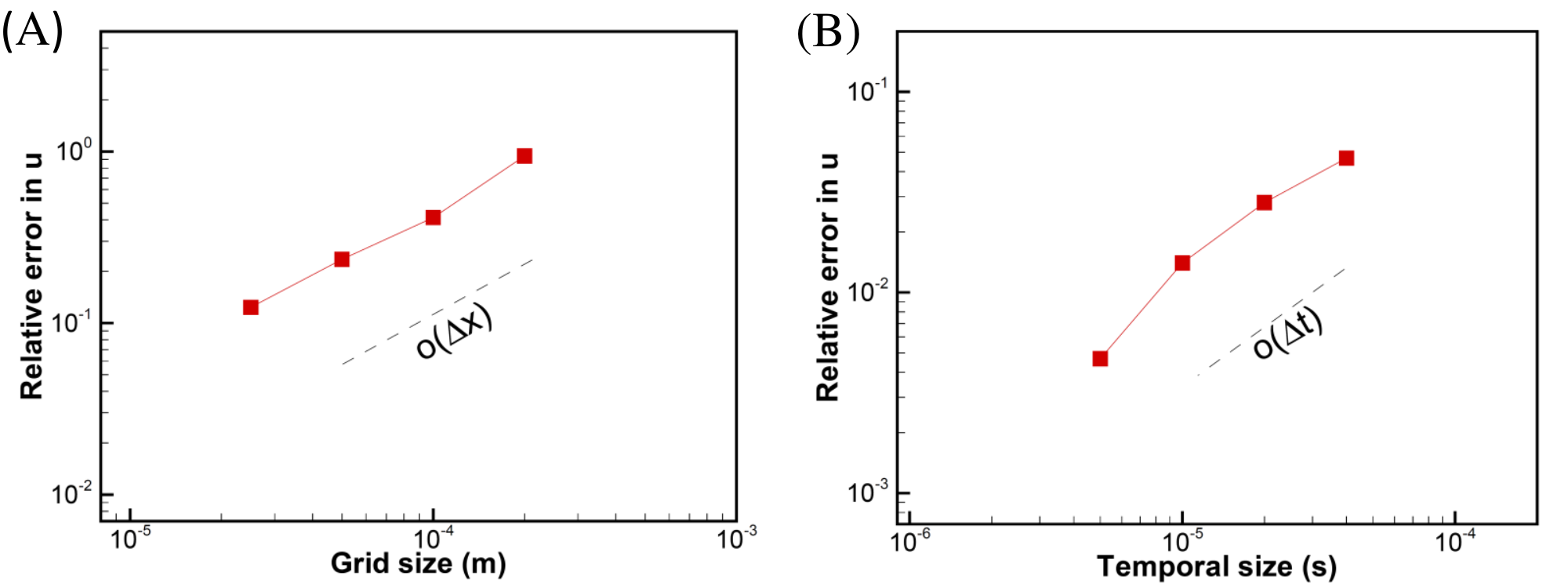}
	\caption{\small Convergence tests for the relative LI error of numerical rise speed. Convergence of the relative error is 1st order for both spatial (\textbf{A}) and temporal (\textbf{B}) resolution. }
	\label{fig:S2}
\end{figure}

In figure \ref{fig:S2}, we reproduce experiment \#5 to evaluate that the simulation converges as the spatial (figure S2A) and temporal (figure S2B) resolution are increased. We test convergence by comparing the numerical rise speed to the corresponding analytical solution. The numerical rise speed is taken along a horizontal cross section in an area of well developed bidirectional flow. Both convergence rates are 1st order. To increase the accuracy of the numerical model efficiently, we apply an adaptive grid refinement strategy. In the simulation of both immiscible and miscible flow, our grid hierarchy is composed of two levels. Level 1 is a fixed, coarse grid that defines the whole computational domain. Level 2 is an adaptively generated, fine grid that overlays the coarse level. The fine grid covers the whole liquid-liquid interface to accurately simulate the evolution of flow regimes. As the bidirectional flow develops and the interface extends through the domain, the area covered by the fine grid is increased as well. Compared to a simulation using the finer grid resolution over the entire computational domain, this adaptive grid refinement strategy obtains the same high accuracy, but saves more than 50\% of the computational expense. More details of our adaptive grid refinement strategy are discussed in \citet{Qin2017}.

%\begin{figure}[t]
%	\centering
%	\includegraphics[width=0.8\textwidth]{../figures/figure_S3/FigS3.pdf}
%	\caption{\small Illustration of our adaptive grid refinement strategy for immiscible and miscbile flow. }
%	\label{fig:AGR}
%\end{figure}

\subsection{Comparison of the average rise speed to the frontal rise speed}

\begin{figure}[t]
	\centering
	\includegraphics[width=0.8\textwidth]{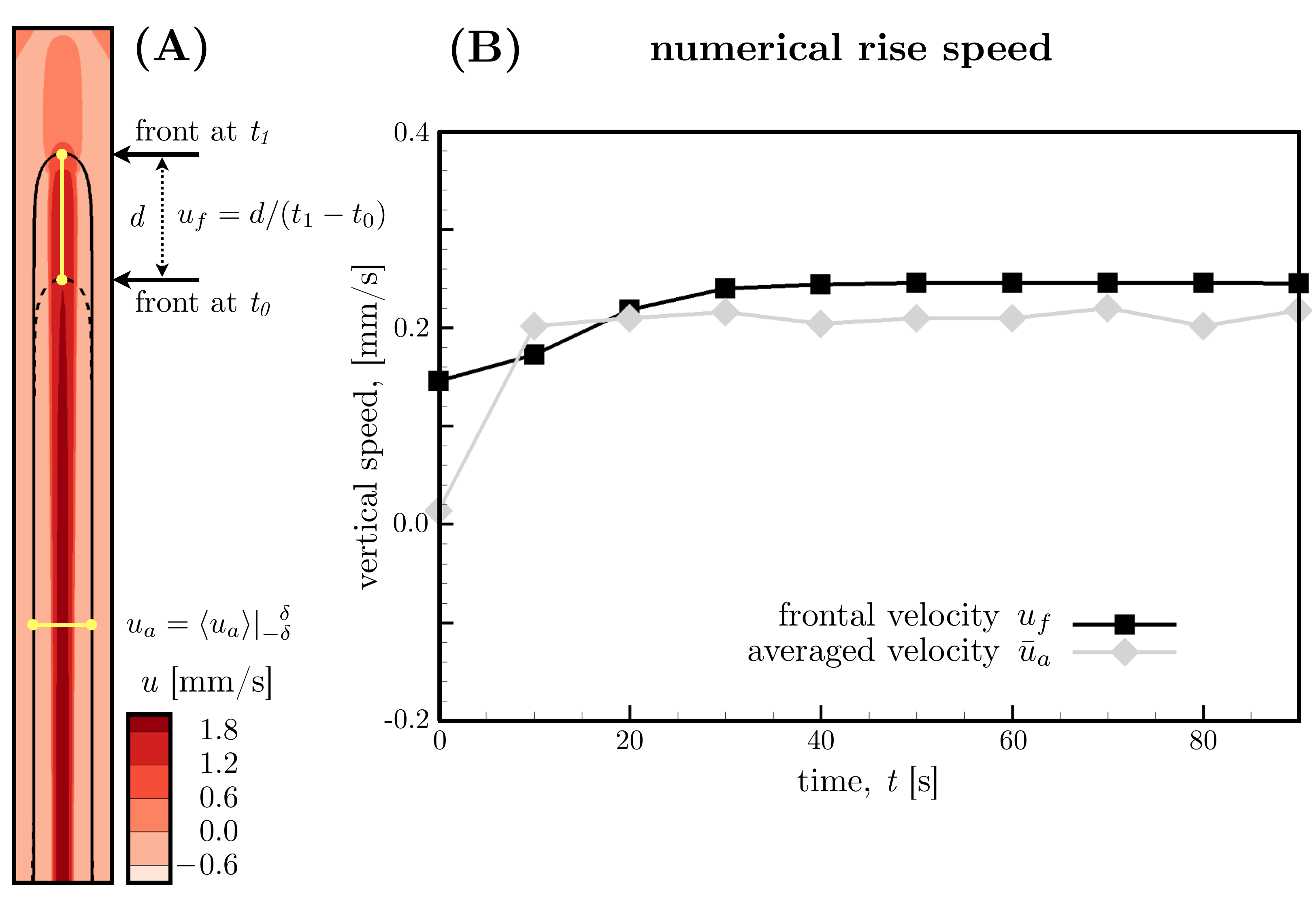}
	\caption{\small Illustration of two numerical metrics for the rise speed in our virtual reproduction of experiment \#8. \textbf{(A)} The frontal rise speed, $u_f$, is computed by measuring the distance of the front of the ascending liquid has covered in a given time interval; the average rise speed, $\bar{u}_a$ is defined by averaging the vertical velocity of the ascending liquid along a horizontal cross section taken in an area of fully developed bidirectional flow. \textbf{(B)} Both $u_f$ and $\bar{u}_a$ reach similar steady state values soon after an initial spin-up period.}
	\label{fig:S3}
\end{figure}

In their analogue experiments, \citet{Stevenson1998} calculated the Ps number using the frontal rise speed of the ascending liquid. Our analytical model assumes that core-annular flow is at steady state, which is not compatible with a propagating front. To compute Ps with our analytical model we use the average rise speed, which we define as the average of the vertical velocity component in horizontal direction across the ascending liquid in the core, and use the experimental Ps numbers to estimate the core radius. Figure \ref{fig:S3} compares the two velocity measures for experiment \#8. Figure \ref{fig:S3}(B) shows that both frontal rise speed and average rise velocity reach a similar steady state after an initial spin-up. We also compare the two rise-speed estimates for other experiments and found that the difference is always small ($<$5\%) and will decrease with spatial and temporal resolution. 

\subsection{Comparison of linear to nonlinear dependence of viscosity on concentration}

\begin{figure}[t]
	\centering
	\includegraphics[width=1\textwidth]{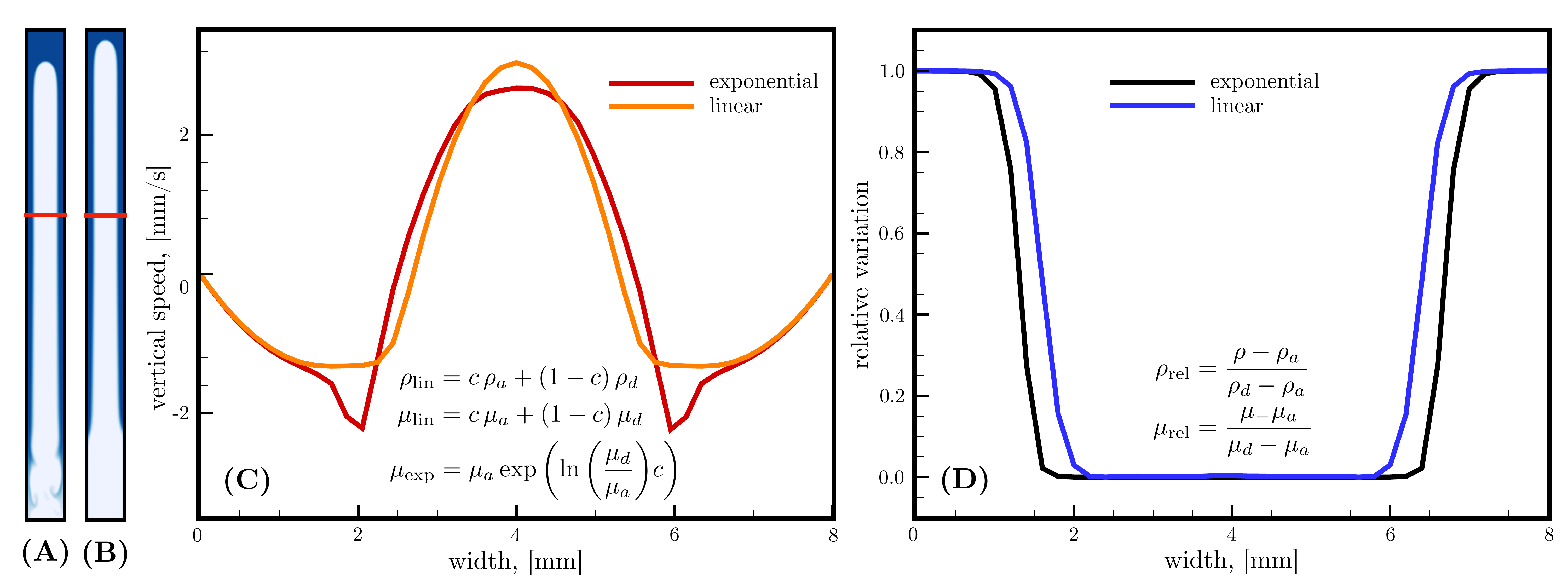}
	\caption{\small Reproduction of experiment \#5 using the linear (A) and exponential (B) dependence of viscosity on concentration. C: Comparison of the numerical velocity profiles between linear (yellow curve) and exponential (red curve) cases  on the cross sections represented by the red line in plot A and B. D: Relative density (blue curve) and viscosity (black curve) profiles of the exponential case on the cross section.}
	\label{fig:S4}
\end{figure}

Several previous numerical studies of the miscible fluids applied nonlinear dependence of viscosity on concentration instead of a linear one used in the present work. For example, \citet{meiburg2004density} defined an exponential relationship between viscosity and concentration as
\begin{equation}
	\mu=\mu_a e^{ln\frac{\mu_d}{\mu_a}c} \ .
\end{equation}
In figure \ref{fig:S4}, we compare the simulation with a linear viscosity profile to the one with an exponential viscosity profile for experiment \#5. In both simulations, we use a linear density profile. We find that the exponential viscosity profile is associated with a smoother flow regime (\ref{fig:S4}B) and a notably different speed profile (\ref{fig:S4}C) as compared to the linear case. The exponential viscosity profile enforces an increased downwelling velocity of the buoyant fluid. Since the density profile is linear, the buoyancy force changes linearly across the interface, whereas the viscous resistance changes exponentially (Figure~\ref{fig:S4}D). That means that there is a thin boundary layer of relatively heavy fluid that has relatively low viscosity and is therefore actively sinking. This actively sinking boundary layer creates the downward velocity peaks shown in figure \ref{fig:S4}C. 

\subsection{Viscous dissipation for fully developed laminar concentric core-annular flow} 

We compute the  dimensionless viscous dissipation  for  fully-developed, laminar, and  concentric core-annular flow by integrating the local viscous dissipation rate over the cross sectional area of the conduit, yielding
\begin{equation}
\Phi=			\int_\delta^1   2 \pi r \left( \frac{du_d}{ d r}\right)^2  dr	+ \frac{1}{M}\int_0^\delta   2 \pi r \left( \frac{du_a}{ d r}\right)^2  dr =		\frac{{\mathrm{Te}}}{\cos \alpha } \ ,
\end{equation}
where $u_d$ and $u_a$ are the velocity profiles of the descending and ascending phase respectively, see eqs. 2.18-19;  the  expression of Transport number  ${\mathrm{Te}}$ is given in eq. 2.25. The viscous dissipation is proportional to the dimensionless flux and, in case of vertical core annular flow ($\cos \alpha =1$), we get 
$\Phi={\mathrm{Te}}$.

\subsection{Interfacial speed for different viscosity contrasts} 

\begin{figure}[t]
	\centering
	\includegraphics[width=0.75\textwidth]{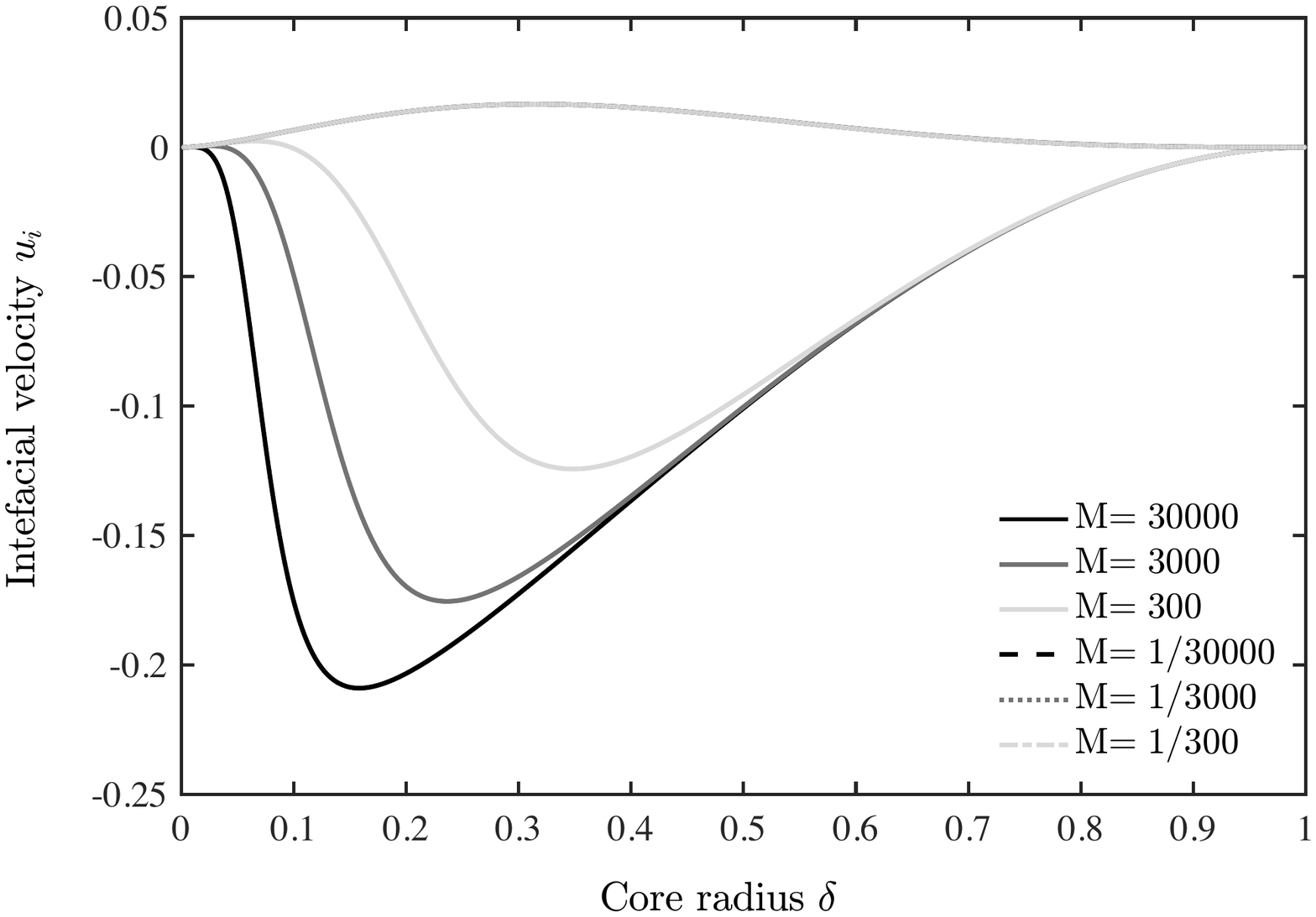}
	\caption{\small The dimensionless interfacial velocity for the core-annular flow configuration as a function of the core-thickness $\delta$ and the viscosity ratio M: negative $u_i$ implies flow reversal in the ascending (core) phase,  positive $u_i$ implies flow reversal in the descending phase. }
	\label{fig:ui}
\end{figure}

The direction of motion of the interface determines the phase in which flow reversal occurs. As evident from out analytical model, the sign of $u_i$ depends sensitively on the driving force, which in turn depends on the viscosity contrast. Figure~\ref{fig:ui} shows the dependence of the dimensionless interfacial speed $u_i$ on the core radius, $\delta$ and the viscosity contrast, M. Negative interface speeds are associated with flow reversal in the ascending fluid and positive interface speeds lead to flow reversal in the descending fluid. When the viscosity of the descending phase is significantly higher than that of the ascending core (${\mathrm{M}} > 10$), flow reversal occurs in the ascending phase almost across the whole range of $\delta$, except for extremely low core radii that are likely dynamically unstable. Conversely, as M decreases, flow reversal begins to shift into the descending phase over a growing range of core radii. If, however, the viscosity ratio is reversed from the experimental setup in \citet{Stevenson1998}, i.e., if the buoyant fluid is now more viscous than the descending fluid, flow reversal occurs in the descending phase across the entire range of $\delta$. 

%\bibliographystyle{jfm}
%\nobibliography{bi-direction}

\end{document}